\documentclass[10pt,aps,prd,onecolumn,showpacs,amsmath,amssymb,nofootinbib,eqsecnum,preprintnumbers,superscriptaddress]{revtex4-2}

\usepackage{mathtools}					
\usepackage{centernot}                  
\usepackage[offset=1.1em]{simpler-wick} 

\usepackage[colorlinks=true,
            citecolor=red,
            linkcolor=blue,
            urlcolor=violet,
            filecolor=cyan,
            backref=false]{hyperref}

\newcommand\bs[1]{\boldsymbol{#1}}

\newcommand\dd{\mathrm{d}}
\newcommand\pp{\partial}
\newcommand\DD{\mathrm{D}}
\newcommand{\nDelta}{{\centernot{\Delta}}}

\newcommand\feq{\mathrel{\phantom{=}}}

\renewcommand{\Re}{\operatorname{Re}}
\renewcommand{\Im}{\operatorname{Im}}
\newcommand{\Ei}{\operatorname{Ei}}
\newcommand{\lie}{\pounds}

\DeclareMathOperator*{\sumint}{%
\mathchoice%
  {\ooalign{$\displaystyle\sum$\cr\hidewidth$\displaystyle\int$\hidewidth\cr}}
  {\ooalign{\raisebox{.14\height}{\scalebox{.7}{$\textstyle\sum$}}\cr\hidewidth$\textstyle\int$\hidewidth\cr}}
  {\ooalign{\raisebox{.2\height}{\scalebox{.6}{$\scriptstyle\sum$}}\cr$\scriptstyle\int$\cr}}
  {\ooalign{\raisebox{.2\height}{\scalebox{.6}{$\scriptstyle\sum$}}\cr$\scriptstyle\int$\cr}}
}

\newcommand\bwpart[1]{{\langle{#1}\rangle}}

\begin{document}


\title{Exact gyratons in higher and infinite derivative gravity }

\author{Ivan Kol\'a\v{r}}
\email{i.kolar@rug.nl}
\affiliation{Van Swinderen Institute, University of Groningen, 9747 AG, Groningen, Netherlands}

\author{Tom\'a\v{s} M\'alek}
\email{malek@math.cas.cz}
\affiliation{Institute of Mathematics of the Czech Academy of Sciences, \v{Z}itn\'a 25, 115 67 Prague 1, Czech Republic}

\author{Suat Dengiz}
\email{sdengiz@thk.edu.tr}
\affiliation{Department of Mechanical Engineering, University of Turkish Aeronautical Association, 06790 Ankara, Turkey}

\author{Ercan Kilicarslan}
\email{ercan.kilicarslan@usak.edu.tr}
\affiliation{Department of Mathematics, Usak University, 64200, Usak, Turkey}

\date{\today}

\begin{abstract}
We study solutions describing spinning null sources called gyratons in generic theories of gravity with terms that are quadratic in curvature and contain an arbitrary number of covariant derivatives. In particular, we show that the properties of pp-waves of the algebraic type III allow for extreme simplification of the field equations. It turns out that the resulting differential equations are exactly solvable due to partial decoupling and linearity of the equations. This is demonstrated explicitly by finding axially symmetric gyraton solutions in Stelle's fourth derivative gravity and the non-local gravity with an infinite number of derivatives.
\end{abstract}

\maketitle


\section{Introduction}

In \textit{general relativity (GR)}, exact interior and exterior solutions of the field equations with a spinning null matter were first studied by Bonnor in 1970 \cite{Bonnor:1970sb}. Later, the gravitational fields generated by spinning ultrarelativistic particles were studied by Frolov and Fursaev \cite{Frolov:2005in}, who called them gyratons. They were further investigated and generalized in \cite{Frolov:2005zq,Frolov:2005ja,Podolsky:2014lpa} (see also \cite{Griffiths:2009dfa}). These geometries belong to the class of pp-wave metrics with non-diagonal terms in the Brinkmann form. Besides the null radiation component, the energy-momentum tensor involves also non-diagonal terms corresponding to internal angular momentum of gyratons. Inside the source, the gyraton spacetimes are of algebraic type III with the aligned null direction being the generator of null geodesics along the direction of propagation. Gyraton solutions beyond the class of type III pp-waves within more general Kundt spacetimes are also known. Namely, gyratons were studied on (anti-)de Sitter background \cite{Frolov:2005ww} (conformal to gyraton pp-waves), Melvin universe \cite{Kadlecova:2010je}, and direct-product spacetimes \cite{Krtous:2012qa}. Lastly, it turns out that gyratons exist not only in non-expanding Kundt geometries, but also in the Robinson--Trautman class of spacetimes~\cite{Podolsky:2018oov}.

In higher derivative theories of gravity, the Einstein--Hilbert action is modified by adding extra scalar curvature invariants. Their field equations are very complex. In order to find their exact solutions, it is often necessary to employ an appropriate ansatz that reduce the equations considerably. A simple ansatz is the so-called \textit{universal spacetimes} \cite{Coley:2008th,Hervik:2013cla,Hervik:2015mja,Hervik:2017sdr}, for which all rank-2 tensors constructed from the metric, the Riemann tensor, and its covariant derivatives are multiples of the metric. The only component of the vacuum field equations then gives an algebraic constraint relating the value of the constant Ricci scalar with the parameters of the given theory. Examples of universal spacetimes of all algebraically special types (II, D, III, N) are known. Generalization of universal spacetimes by relaxing the condition imposed on the rank-2 tensors lead to the \textit{almost universal spacetimes} \cite{Kuchynka:2018ezs}; the rank-2 tensors allowed have a tracefree part of type N. Therefore, the field equations for almost universal spacetimes are compatible with energy-momentum tensors of null radiation and reduce to the algebraic constraint accompanied by a single partial differential equation.

The possible scalar invariant quadratic in curvature (without additional derivative) that can be added to the action are Ricci scalar square, Ricci tensor square, and Riemann tensor square. A generic theory that contains all three is called the \textit{Stelle gravity (SG)} \cite{Stelle:1977ry}. Some exact solutions of this theory were described in \cite{Malek:2011pn,Gurses:2012db,Pravda:2016fue}. More complicated geometry ansatz (such as the spherically symmetric spacetime) usually requires numerical treatment or the use of the Frobenius method to find infinite series solution \cite{Lu:2015cqa,Lu:2015psa,Podolsky:2019gro}. 

Going one step further, one can consider quadratic terms that also contain covariant derivatives. A particularly interesting theory arises when one takes an infinite number of derivatives, the so-called \textit{infinite derivative gravity (IDG)} \cite{Tomboulis:1997gg,Modesto:2011kw,Biswas:2011ar} (discussed already in \cite{Krasnikov:1987yj}). This \textit{non-local} theory attracted attention for its predisposition for resolving singularities together with retaining the same degrees of freedom (around Minkowski background) as there are in GR. Unfortunately, this theory leads to immensely difficult non-local non-linear field equations, for which the numerical methods are not available, so the research has focused mainly on solutions in the linearized (weak-field) regime of the theory \cite{Frolov:2015usa,Frolov:2015bia,Buoninfante:2018stt,Boos:2018bxf,Buoninfante:2018xif,Kolar:2020bpo,Kolar:2021oba}. Gyratons in linearized IDG were constructed in \cite{Boos:2020ccj} by boosting a solution for stationary spinning object and taking its Penrose limit \cite{Penrose1976} in an analogy to the Aichelburg--Sexl ultraboost of a point-like source \cite{Aichelburg1971}. An advancement in the search for exact solutions of the full IDG came with the almost universal spacetimes \cite{Kolar:2021rfl}, which explained the previous discovery of exact shock/impulsive waves in IDG \cite{Lousto:1996ep,Kilicarslan:2019njc,Dengiz:2020xbu}.\footnote{Other exact solutions were obtained in the context of cosmology, where the field equations were effectively localized by imposing a recurrent ansatz on Ricci scalar \cite{Biswas:2005qr,Biswas:2010zk,Biswas:2012bp,Koshelev:2017tvv,Kumar:2020xsl}.} With this ansatz, the field equations effectively reduce to a linear but still non-local partial differential equation, which can be solved exactly by eigenfunction expansion or using the heat kernel method. The obtained solutions represent gravitational waves generated by null particles propagating in maximally symmetric spacetimes.

In this paper, we step further out of the family of almost universal spacetimes by permitting the rank-2 tensors to have a tracefree part of type III. We show that many terms in the field equations of theories of gravity with an arbitrary number of covariant derivatives either vanish or simplify significantly for type III pp-waves. This allows us to find the reduced field equations for gyraton metric ansatz, which are compatible with energy-momentum tensors of spinning null sources. These equations can be solved exactly as we show by finding exact axially symmetric gyraton solutions in full Stelle gravity as well as full infinite derivative gravity.

The paper is structured as follows: In Section \ref{sec:genquadcurvgrav} we introduce the action and the field equations of a generic gravity that is quadratic in curvature with arbitrary analytic (covariant) differential operators. In Section \ref{sec:typeIIIppwaves} we examine various properties of the pp-waves of algebraic type III. In particular, we show that many terms that appear in the field equations either vanish or simplify significantly. In Section \ref{sec:feqgyr}, we derive the field equations for gyratons and demonstrate that they are exactly solvable thanks to partial decoupling and linearity. Sections \ref{sec:gyrgr}, \ref{sec:gyrsg}, and \ref{sec:gyridg} are devoted to specific examples. We review and find new solutions in the general relativity, the Stelle gravity, and the infinite derivative gravity. In section \ref{sec:highordcurv}, we comment on gravity theories of higher order in curvature. The paper is concluded with a brief discussion of our results in Section \ref{sec:C}. Appendices \ref{apx:NP}, \ref{apx:explcalcSS}, and \ref{apx:auxstattypeIIIpp} provide supplementary material.

\subsection*{Index/index-free tensor notation}

Before we proceed to the main calculations, let us clarify the tensor notation employed in this paper. We use the bold font for tensors and their indices, which are understood as the \textit{abstract tensor indices} \cite{penrose_rindler_1984} $\bs{a}, \bs{b}, \dots$, thus indicating tensor type rather than their components. The regular font is used for scalar quantities (such as coordinates and tensor components) and non-tensorial indices. On top of that, we also employ the index-free notation, where $\cdot$ indicates the contraction between two adjacent tensor indices. For example, the contraction of a vector $\bs{v}$ with a covector $\bs{c}$ reads $\bs{v}\cdot\bs{c}=\bs{v^a c_a}$. Raising and lowering of all tensor indices is achieved with the help of the \textit{musical isomorphisms} $^\sharp$ and $^\flat$ \cite{Lee2012}; e.g., ${\bs{c}^{\sharp}=\bs{g}^{-1}\cdot\bs{c}}$ and ${\bs{v}^{\flat}=\bs{g}\cdot\bs{v}}$. Furthermore, we use $\vee$ and $\wedge$ to denote the symmetric and anti-symmetric tensor products, ${\bs{p}\vee\bs{q}=\bs{p}\bs{q}+\bs{q}\bs{p}}$ and ${\bs{p}\wedge\bs{q}=\bs{p}\bs{q}-\bs{q}\bs{p}}$.


\section{Generic quadratic curvature gravity}\label{sec:genquadcurvgrav}
Consider a 4-dimensional manifold $M$ equipped with a metric $\bs{g}$. A generic gravity action that is quadratic in the Riemann tensor $\bs{R}$ and analytic in the covariant derivative $\bs{\nabla}$ can be written as
\begin{equation}
    S=\frac12\int_M \! \mathfrak{g}^{\frac12} \, \big[\varkappa^{-1}\big(R-2\Lambda\big)+\bs{R^{abcd}}\bs{\mathcal{D}(\bs{\nabla}){}_{abcd}{}^{efgh}}\bs{R_{efgh}}\big]+S_{\textrm{m}}\;,
\end{equation}
where ${\bs{\mathcal{D}}}$ stands for an arbitrary tensorial-differential operator. Using the symmetries of the Riemann tensor and the Bianchi identities, one can show that, up to the higher-order terms in curvature, the action can be recast to the form~\cite{Biswas:2011ar,Biswas:2013kla}
\begin{equation}\label{eq:action}
    S\sim\frac12\int_M \! \mathfrak{g}^{\frac12} \, \big[\varkappa^{-1}\big(R-2\Lambda\big)+R\mathcal{F}_1(\square)R+\bs{S^{ab}}\mathcal{F}_2(\square)\bs{S_{ab}}+\bs{C^{abcd}}\mathcal{F}_3(\square)\bs{C_{abcd}}\big]+S_{\textrm{m}}\;,
\end{equation}
where $R$ is the Ricci scalar, $\bs{S_{ab}}$ is the tracefree (TF) Ricci tensor, $\bs{C_{abcd}}$ is the Weyl tensor, and $\mathfrak{g}^{\frac12}$ denotes the volume element (1-density associated to the metric $\bs{g}$). The action \eqref{eq:action} contains the \textit{form-factors}, which are analytic functions of the wave operator ${\square\equiv\bs{\nabla}^{\sharp}\cdot\bs{\nabla}}$,
\begin{equation}
    \mathcal{F}_{i}(\square)=\sum_{n=0}^{\infty}f_{i,n}\square^n\;,
\end{equation}
where $f_{i,n}$ are arbitrary constant coefficients. Varying the action \eqref{eq:action} with respect to the metric $\bs{g}$, one can find \cite{Biswas:2013cha}\footnote{Contrary to \cite{Biswas:2013cha}, we added the missing symmetrization to the terms that are not symmetric (in general) in indices $\bs{a}$, $\bs{b}$ and used the identity $\bs{\bs{C}_{(\bs{a}}{}^{\bs{cde}}} \square^n \bs{\bs{C}_{\bs{b})\bs{cde}}} = \frac14 \bs{g_{ab} C^{cdef}} \square^k \bs{C_{cdef}}$.
This identity follows from theorem 3(a) in \cite{Lovelock_1970} for the Weyl tensor in four dimensions, ${\bs{\bs{C}_{[\bs{cd}}{}^{[\bs{ef}} \bs{\delta}_{\bs{b}]}{}^{\bs{a}]}} = 0}$,
when contracted with $\square^n \bs{C_{ef}{}^{cd}}$.} 
\begin{equation}\label{eq:fieldeq}
\begin{aligned}
    &\varkappa^{-1}\big(\bs{S_{ab}}-\tfrac14 R\bs{g_{ab}}+\Lambda\bs{g_{ab}}\big)+2\bs{S_{ab}}\mathcal{F}_1(\square) R -2 \big(\bs{\nabla_a\nabla_b} -\bs{g_{ab}}\square\big) \mathcal{F}_1(\square) R +\big(\square+\tfrac12 R\big)\mathcal{F}_2(\square) \bs{S_{ab}}
    \\
    &-2\bs{g_{d(a}}\big(\bs{\nabla^c}\bs{\nabla^{d}}-\bs{S^{cd}}\big) \mathcal{F}_2(\square) \bs{S_{b)c}} +\bs{g_{ab}} \big(\bs{\nabla^c} \bs{\nabla^d}-\tfrac12\bs{S^{cd}}\big) \mathcal{F}_2(\square) \bs{S_{cd}}-4\big(\bs{\nabla^c}\bs{\nabla^d}+\tfrac12\bs{S^{cd}}\big)\mathcal{F}_3(\square)\bs{C_{d(ab)c}}
    \\
    &-\bs{\Omega}_1\bs{{}_{ab}}+\tfrac12 \bs{g_{ab}}\big(\bs{\Omega}_1\bs{{}^c{}_c}+\mho_1\big)-\bs{\Omega}_2\bs{{}_{ab}}+\tfrac12 \bs{g_{ab}}\big(\bs{\Omega}_2\bs{{}^c{}_c}+\mho_2\big)-\bs{\Omega}_3\bs{_{ab}}+\tfrac12 \bs{g_{ab}}\big(\bs{\Omega}_3\bs{{}^c{}_c}+\mho_3\big)-2\bs{\Upsilon}_2\bs{{}_{ab}}-4\bs{\Upsilon}_3\bs{{}_{ab}}=\bs{T_{ab}}\;,
\end{aligned}
\end{equation}
where the symmetric tensors $\bs{\Omega}_i\bs{{}_{ab}}$, $\bs{\Upsilon}_i\bs{{}_{ab}}$, and scalars ${\mho_i}$ are given by double-sums:
\begin{equation}\label{eq:LK}
\begin{gathered}
\begin{aligned}
    \bs{\Omega}_1\bs{{}_{ab}} &=\sum_{n=1}^\infty f_{1,n}\sum_{l=0}^{n-1}\bs{\nabla_a} \square^l R\,\bs{\nabla_b} \square^{n-l-1}R\;, & \mho_1 &=\sum_{n=1}^\infty f_{1,n}\sum_{l=0}^{n-1}\square^l R \,\square^{n-l}R\;,
    \\
    \bs{\Omega}_2\bs{{}_{ab}} &=\sum_{n=1}^\infty f_{2,n}\sum_{l=0}^{n-1}\bs{\nabla_a} \square^l \bs{S^{cd}}\,\bs{\nabla_b} \square^{n-l-1}\bs{S_{cd}}\;, & \mho_2 &=\sum_{n=1}^\infty f_{2,n}\sum_{l=0}^{n-1}\square^l \bs{S^{cd}} \,\square^{n-l}\bs{S_{cd}}\;,
    \\
    \bs{\Omega}_3\bs{{}_{ab}} &=\sum_{n=1}^\infty f_{3,n}\sum_{l=0}^{n-1}\bs{\nabla_a} \square^l \bs{C^{cdef}}\,\bs{\nabla_b} \square^{n-l-1}\bs{C_{cdef}}\;, & \mho_3 &=\sum_{n=1}^\infty f_{3,n}\sum_{l=0}^{n-1}\square^l \bs{C^{cdef}} \,\square^{n-l}\bs{C_{cdef}}\;,
\end{aligned}
\\
\begin{aligned}
    \bs{\Upsilon}_2\bs{{}_{ab}} &=\sum_{n=1}^\infty f_{2,n}\sum_{l=0}^{n-1} \bs{\nabla_c}\big[\square^l \bs{S^{cd}}\,\bs{\nabla_{(a}} \square^{n-l-1}\bs{S_{{b)}d}}-\bs{\nabla_{(a}}\square^l \bs{S^{cd}} \,\square^{n-l-1}\bs{S_{{b)}d}}\big]\;,
    \\
    \bs{\Upsilon}_3\bs{{}_{ab}} &=\sum_{n=1}^\infty f_{3,n}\sum_{l=0}^{n-1} \bs{\nabla_c}\big[\square^l \bs{C^{cdef}}\,\bs{\nabla_{(a}} \square^{n-l-1}\bs{C_{{b)}def}}-\bs{\nabla_{(a}}\square^l \bs{C^{cdef}}\, \square^{n-l-1}\bs{C_{{b)}def}}\big]\;.
\end{aligned}
\end{gathered}
\end{equation}


\section{Type III pp-waves}\label{sec:typeIIIppwaves}
\subsection{Generic pp-waves}
The class of \textit{pp-wave} metrics describing plane-fronted waves with parallel rays is geometrically defined by the property that it admits a \textit{covariantly constant null vector (CCNV)} $\bs{\xi^a}$, 
\begin{equation}\label{eq:CCNVl}
    \bs{\nabla} \bs{\xi} = 0\;.
\end{equation}
It immediately follows that $\bs{\xi}$ is a Killing vector, ${\lie_{\bs{\xi}} \bs{g} = 0}$, and consequently $\bs{\xi}$ is a curvature collineation,
\begin{equation}\label{eq:CCNVDR}
    \lie_{\bs{\xi}} \bs{R} = \bs{\nabla_\xi R} = 0\;,
\end{equation}
where $\bs{\nabla_\xi} \equiv \bs{\xi} \cdot \bs{\nabla}$. On the other hand, the Ricci identities for CCNV $\bs{\xi}$ \eqref{eq:CCNVl} imply
\begin{equation}\label{eq:CCNVl*R}
    \bs{\xi} \cdot \bs{R} = 0\;,
\end{equation}
and therefore $\bs{\nabla}$ and $\bs{\nabla_\xi}$ commute when applied on an arbitrary tensor $\bs{t}$,
\begin{equation}
    [\bs{\nabla_\xi},\bs{\nabla}]\bs{t} = 0\;.
\end{equation}
Substituting the Riemann tensor $\bs{R}$ for $\bs{t}$, it turns out that
\begin{equation}\label{eq:DnablaR}
    \bs{\nabla_\xi \nabla}{\cdots}\bs{\nabla R} = 0\;.
\end{equation}

Throughout the paper, we will make extensive use of the \textit{Newman--Penrose (NP) formalism}, which is summarized in Appendix \ref{apx:NP}. It utilizes the \textit{orthonormal null covector frame} $\{\bs{e}^i\}_{i=0,\ldots,3}$ consisting of two real null covectors $\bs{e}^0_{\bs{a}} \equiv \bs{l_a}$, $\bs{e}^1_{\bs{a}} \equiv \bs{n_a}$, a complex null covector $\bs{e}^2_{\bs{a}} \equiv \bs{m_a}$, and its conjugate $\bs{e}^3_{\bs{a}} \equiv \bar{\bs{m}}{}_{\bs{a}}$ satisfying
\begin{equation}\label{eq:orthonormalframe}
    \bs{l}^{\sharp} \cdot \bs{n} = -1\;,
    \quad
    \bs{m}^{\sharp} \cdot \bar{\bs{m}} = 1\;,
\end{equation}
where, for convenience, we identify the CCNV $\bs{\xi}$ of pp-waves with the null frame vector ${\bs{l}^{\sharp}=\bs{\xi}}$. The metric and its inverse can be then written as
\begin{equation}\label{eq:framemetric}
    \bs{g} =-\bs{l}\vee\bs{n}+\bs{m}\vee\bar{\bs{m}}\;,
    \quad
    \bs{g}^{-1} =-\bs{l}^{\sharp}\vee\bs{n}^{\sharp}+\bs{m}^{\sharp}\vee\bar{\bs{m}}^{\sharp}\;.
\end{equation}
Various contractions of \eqref{eq:CCNVl*R} with appropriate frame vectors lead to the following relations of the frame components of the TF Ricci and the Weyl tensor (defined in \eqref{eq:NPformalismCS}):
\begin{equation}\label{eq:CCNVPhiPsi}
    \Psi_0 = \Psi_1 = \Phi_{00} = \Phi_{01} = 0, \quad 3 \Psi_2 = 2 \Phi_{11} = - R/4\;, \quad \Psi_3 = - \Phi_{21}\;.
\end{equation}
Therefore, the TF Ricci tensor and the Weyl tensor of generic pp-wave metrics are both of the algebraic types~II and specialize to types~III if the Ricci scalar vanishes (see, e.g., \cite{Ortaggio:2012jd} for a review of algebraic classification based on null alignment which is equivalent to Petrov classification of the Weyl tensor in four dimensions).

\subsection{Type III}\label{ssec:typeIIIppwaves}

Let us focus on the class of pp-wave spacetimes for which the TF Ricci tensor and Weyl tensor are both of the algebraic types~III. This means that there exists an \textit{aligned null direction} $\bs{l}^{\sharp}$ such that $\bs{S}$ and $\bs{C}$ contain the following components only:\footnote{To familiarize the reader with our notation, we write \eqref{eq:SCtypeIII} also using tensor indices,
\begin{equation*}
\begin{aligned}
    \bs{S_{ab}} &=-4\Phi_{21} \bs{l_{(a}m_{b)}}-4\bar{\Phi}_{21} \bs{l_{(a}\bar{m}_{b)}}+2\Phi_{22} \bs{l_a l_b}\;,
    \\
    \bs{C_{abcd}} &=-8\Psi_3 \bs{l_{\{a}n_{b}l_{c}m_{d\}}}-8\bar{\Psi}_3 \bs{l_{\{a}n_{b}l_{c}\bar{m}_{d\}}} - 8\Psi_3 \bs{l_{\{a}m_{b}\bar{m}_{c}m_{d\}}} - 8\bar{\Psi}_3 \bs{l_{\{a}\bar{m}_{b}m_{c}\bar{m}_{d\}}}
    \\
    &\feq+4 \Psi_4 \bs{l_{\{a}m_{b}l_{c}m_{d\}}} + 4 \bar{\Psi}_4 \bs{l_{\{a}\bar{m}_{b}l_{c}\bar{m}_{d\}}}\;.
\end{aligned}
\end{equation*}
where the curly brackets correspond to ${\bs{X_{\{abcd\}}}\equiv\tfrac12(\bs{X_{[ab][cd]}}+\bs{X_{[cd][ab]}})}$.}
\begin{equation}\label{eq:SCtypeIII}
\begin{aligned}
    \bs{S} &= -2 \bs{l}\vee\big(\Phi_{21}\bs{m}+ \bar{\Phi}_{21}\bar{\bs{m}}\big)+ 2\Phi_{22}\,\bs{l}\bs{l}\;,
    \\
    \bs{C} &=\Psi_3 (\bs{l}\wedge\bs{m})\vee(\bs{n}\wedge\bs{l}+\bs{m}\wedge\bar{\bs{m}})+\bar{\Psi}_3 (\bs{l}\wedge\bar{\bs{m}})\vee(\bs{n}\wedge\bs{l}+\bar{\bs{m}}\wedge\bs{m})
    \\
    &\feq+\Psi_4(\bs{l}\wedge\bs{m})(\bs{l}\wedge\bs{m})+\bar{\Psi}_4(\bs{l}\wedge\bar{\bs{m}})(\bs{l}\wedge\bar{\bs{m}})\;.
\end{aligned}
\end{equation}
Then ${R = 0}$ due to \eqref{eq:CCNVPhiPsi}.
On top of that, we require that the null frame is \textit{parallel-propagated (PP)} along $\bs{l}^{\sharp}$,
\begin{equation}\label{eq:Dnm}
    \DD\bs{n} = \DD\bs{m} = \DD\bar{\bs{m}} = 0\;,
\end{equation}
where $\DD\equiv\bs{l}^{\sharp}\cdot\bs{\nabla}$.
Inverting the relations \eqref{eq:NPspincoeffs} for the spin coefficients and using \eqref{eq:CCNVl} and \eqref{eq:Dnm}, one can find for the pp-wave geometries in the PP frame,
\begin{equation}\label{eq:spincoeffCCNVppframe}
    \kappa = \tau = \sigma = \rho = \pi = \varepsilon = \gamma + \bar\gamma = \alpha + \bar\beta = 0\;.
\end{equation}
As a consequence of these relations and \eqref{eq:commutonscal}, the directional derivatives $\nDelta\equiv\bs{n}^{\sharp}\cdot\bs{\nabla}$, $\delta\equiv\bs{m}^{\sharp}\cdot\bs{\nabla}$, and $\bar\delta\equiv\bar{\bs{m}}^{\sharp}\cdot\bs{\nabla}$ commute with $\DD$ when acting on scalars,
\begin{equation}\label{eq:Dcomppwave}
    [\nDelta,\DD] =[\delta,\DD] =[\bar\delta,\DD] =0\;.
\end{equation}

Let us discuss the reduction of the field equations of a generic theory for these geometries. The pp-waves of type III are the so-called T-III spacetimes (see Proposition 16 in \cite{Kuchynka:2018ezs}), for which any rank-2 tensor $\bs{B_{ab}}$ constructed from the $\bs{R}$ and $\bs{\nabla}$ of an arbitrary order takes the form
\begin{equation}\label{eq:TIIIspacetimes}
    \bs{B}=\zeta\, \bs{g} + \bar{\psi}\, \bs{l}\vee\bs{m} + \psi\, \bs{l}\vee\bar{\bs{m}} + \omega\, \bs{l} \bs{l}\;.
\end{equation}
with ${\zeta = 0}$ in our case (${R=0}$), since these spacetimes are of \textit{vanishing scalar invariants (VSI)} \cite{Coley:2004hu}. Therefore the field equations for the pp-waves of type III reduce to a system of three partial differential equations (one component $\omega$ and two components of complex $\psi$). Before we look at these components in more detail, we need to introduce several convenient notions. The frame normalization \eqref{eq:orthonormalframe} and therefore the form of the metric \eqref{eq:framemetric} is preserved by the Lorentz transformations of the frame, namely spatial rotations, null rotations and boosts
\begin{equation}
    \bs{l} \rightarrow e^\varpi \bs{l}\;,
    \quad
    \bs{n} \rightarrow e^{-\varpi} \bs{n}\;,
\end{equation}
with a real parameter $\varpi$. We say that a quantity $q$ has a \textit{boost weight} (b.w.) $b$ if it transforms under boosts according~to
\begin{equation}
    q \rightarrow e^{b \varpi}q\;. 
\end{equation}
The \textit{boost order (b.o.)} of a tensor with respect to a given frame is defined as the maximal b.w.\ of its frame components. For example, $\Phi_{21}$ and $\Phi_{22}$ are of b.w.\ $-1$ and $-2$, respectively, as can be seen directly from \eqref{eq:NPformalismCS}. Then $\bs{S}$ in \eqref{eq:SCtypeIII} is obviously of b.o.\ $-1$. We also adopt the balanced scalar approach of \cite{Pravda:2002us} (see also \cite{Hervik:2013cla,Hervik:2015mja,Hervik:2017sdr,Kuchynka:2018ezs}).
In a PP frame along affinely parameterized null geodesics generated by vector field $\bs{l}^{\sharp}$, a tensor $\bs{t}$ is said to be \textit{$k$-balanced}, if its boost weight $b$ part $\bs{t}^\bwpart{b}$ satisfies $\bs{t}^\bwpart{b} = 0$ for $b \geq -k$ and $\DD^{-b-k}\bs{t}^\bwpart{b} = 0$ for $b < -k$. If $\bs{t}$ is 0-balanced, we say it is \textit{balanced}.

In a PP frame \eqref{eq:Dnm} with CCNV $\bs{l}^{\sharp}$, frame components of a tensor $\bs{t} = t_{i_1 \cdots i_p}{}^{j_1 \cdots j_q} \bs{e}^{i_1} \cdots \bs{e}^{i_p} \bs{e}_{j_1} \cdots \bs{e}_{j_q}$ do not change along the null geodesics with $\bs{l}^{\sharp}$ being the generator, i.e.\ $\DD t_{i_1 \cdots i_p}{}^{j_1 \cdots j_q}=0$, if $\DD\bs{t} = 0$ and vice versa. Therefore, \eqref{eq:CCNVDR} ensures that all components of the curvature tensors are annihilated by the operator $\DD$ (this can be seen also from the Bianchi identities \eqref{eq:BianchiIds})
\begin{equation}\label{eq:CCNVBianchi}
\begin{aligned}
    b &= -1: &  \DD\Psi_3 &= \DD\Phi_{21} = 0\;,
    \\
    b &= -2: &  \DD\Psi_4 &= \DD\Phi_{22} = 0\;,
\end{aligned}
\end{equation}
where we also indicate the corresponding boost weight $b$ of given components. The curvature tensors $\bs{S}$ and $\bs{C}$ (and consequently the Riemann tensor $\bs{R}$) of pp-waves of type III are thus balanced. Moreover, b.w.\ $-2$ parts of $\bs{S}$ and $\bs{C}$,
\begin{equation}
\begin{aligned}
    \bs{S}^{\bwpart{-2}} &= 2\Phi_{22}\,\bs{l}\bs{l}\;, \\
    \bs{C}^{\bwpart{-2}} &=\Psi_4(\bs{l}\wedge\bs{m})(\bs{l}\wedge\bs{m})+\bar{\Psi}_4(\bs{l}\wedge\bar{\bs{m}})(\bs{l}\wedge\bar{\bs{m}})\;,
\end{aligned}    
\end{equation}
are 1-balanced.

In general, the covariant derivative of a tensor can increase its boost order. For example, applying the covariant derivative on a rank-$k$ contravariant tensor
$\bs{t} = t^{i_1 \cdots i_k} \bs{e}_{i_1} \cdots \bs{e}_{i_k}$ leads to
$\bs{\nabla}\bs{t}= (\bs{\nabla}t^{i_1 \cdots i_k}) \bs{e}_{i_1} \cdots \bs{e}_{i_k} + t^{i_1 \cdots i_k} (\bs{\nabla}\bs{e}_{i_1}) \cdots \bs{e}_{i_k} + \ldots + t^{i_1 \cdots i_k} \bs{e}_{i_1} \cdots (\bs{\nabla}\bs{e}_{i_k})$. Using the frame decomposition of the covariant derivative \eqref{eq:nablaDDeltadeltadelta}, there appear terms $(\DD t_{i_1 \cdots i_k}) \bs{n} \bs{e}^{i_1} \cdots \bs{e}^{i_k}$ and thus the operator $\DD$ increases boost weights of given components by one. The terms $\bs{\nabla}\bs{e}_i$ in principle add $+2$ or $+1$ to b.w. of the resulting components via the spin coefficients $\kappa$ or $\rho$, $\sigma$, $\varepsilon$, respectively. However, these spin coefficients vanish for pp-waves \eqref{eq:spincoeffCCNVppframe}. Although the possible non-vanishing spin coefficients introduced by $\bs{\nabla}\bs{e}_i$ either do not change or decrease b.w., subsequent application of one more $\bs{\nabla}$ on $\bs{\nabla}\bs{t}$ produces also $\bs{\nabla}$ of these spin coefficients and again due to the decomposition \eqref{eq:nablaDDeltadeltadelta} the operator $\DD$ increases b.w. For instance, the spin coefficient $\alpha$ has b.w.\ 0 and thus $\DD^n\alpha$ is of b.w.\ $n$. Therefore, after several applications of the covariant derivative, there could appear a component of the boost weight exceeding the boost order of the original tensor. Nevertheless, such a situation does not happen in the case of type III pp-waves since the Ricci identities \eqref{eq:RicciIds} for the non-vanishing spin coefficients imply
\begin{equation}
\begin{aligned}
    b &= 0: &  \DD\alpha = \DD\beta = 0\;, 
    \\
    b &= -1: &  \DD\gamma = \DD\lambda = \DD\mu = 0\;, \\
    b &= -2: &  \DD\nu = 0\;.
\end{aligned}
\end{equation}
Similar reasoning carried out formally results in Lemma 1 of \cite{Kuchynka:2018ezs}: the covariant derivative of a $k$-balanced tensor in a degenerate\footnote{The Kundt class is defined geometrically as spacetimes admiting a non-expanding, non-shearing and non-twisting null geodesic congruence. The pp-wave metrics thus belong to a special subfamily of the Kundt class. A Kundt spacetime is said to be degenerate if the Riemann tensor and all its covariant derivatives are algebraically special (i.e.\ of type II or more special) with the generator of the Kundt null geodesic congruence being their aligned null direction.} Kundt spacetime is again a $k$-balanced tensor. One can thus conclude that $\bs{\nabla}{\cdots}\bs{\nabla}\bs{S}$, $\bs{\nabla}{\cdots}\bs{\nabla}\bs{C}$ are balanced and $\bs{\nabla}{\cdots}\bs{\nabla}\bs{S}^\bwpart{-2}$, $\bs{\nabla}{\cdots}\bs{\nabla}\bs{C}^\bwpart{-2}$ are 1-balanced.

The balance property of the curvature tensors has several direct consequences for rank-2 tensors and thus for the field equations. First, recall that k-balanced tensors are of b.o.\ ${-(k{+}1)}$ (i.e.\ $\bs{\nabla}{\cdots}\bs{\nabla}\bs{S}$, $\bs{\nabla}{\cdots}\bs{\nabla}\bs{C}$ are of b.o.\ $-1$ while $\bs{\nabla}{\cdots}\bs{\nabla}\bs{S}^\bwpart{-2}$, $\bs{\nabla}{\cdots}\bs{\nabla}\bs{C}^\bwpart{-2}$ are of b.o.\ $-2$) and rank-2 tensors only admit components with b.w. ranging from $-2$ to $2$. 
Therefore, rank-2 tensors cubic or of a higher order in curvature vanish (a tensor of order $k$ in curvature tensors which are of b.o.\ $-1$ is thus of b.o.\ $-k$). Only b.w.\ $-1$ parts $\bs{S}^\bwpart{-1}$, $\bs{C}^\bwpart{-1}$ and their covariant derivatives contribute to rank-2 tensors quadratic in curvature specifically to the $\omega$ term of \eqref{eq:TIIIspacetimes}. Also, we immediately see that the pp-wave spacetimes of type III are of VSI because all scalars are constructed as contractions of balanced tensors $\bs{\nabla}{\cdots}\bs{\nabla}\bs{S}$ and $\bs{\nabla}{\cdots}\bs{\nabla}\bs{C}$, which are of b.o.\ $-1$.

\subsection{Vanishing tensors quadratic in curvature}
\label{ssec:TensorsQuadraticInCurvature}

In this subsection, we consider tensors (of any rank) that are quadratic in curvature, namely tensors constructed as contractions of $\bs{\nabla}{\cdots}\bs{\nabla}\bs{S}\bs{\nabla}{\cdots}\bs{\nabla}\bs{S}$, $\bs{\nabla}{\cdots}\bs{\nabla}\bs{S}\bs{\nabla}{\cdots}\bs{\nabla}\bs{C}$, or $\bs{\nabla}{\cdots}\bs{\nabla}\bs{C}\bs{\nabla}{\cdots}\bs{\nabla}\bs{C}$. We show that such tensors with specific configurations of indices vanish for type III pp-wave spacetimes:

\begin{itemize}
    \item $\bs{\nabla}{\cdots}\bs{\nabla} \bs{S}\bs{\nabla}{\cdots}\bs{\nabla}\bs{S}$ vanishes if at least one $\bs{S}$ has no free index.

        Without loss of generality we assume that ${\bs{S}}$ with no free index is the first one. Since ${\bs{S}}$ is of type III, the first $\bs{S}$ contains at least one contracted $\bs{l}$, see \eqref{eq:SCtypeIII}. Here, we only sketch the proof using a schematic notation, where $\bs{l}$-contractions are denoted by lines between the contracted expressions (similar to the well-known notation for Wick contractions). Explicit calculations of all combinations are listed in Appendix \ref{apx:explcalcSS}. Null covector $\bs{l}$ from the first $\bs{S}$ may contract in four different ways:
        \begin{equation}\label{eq:SSterms}
        \begin{aligned}
            \bs{\nabla}{\cdots}\bs{\nabla} \overset{\sqcap}{\bs{S}}\bs{\nabla}{\cdots}\bs{\nabla}\bs{S} &=0\;, \\
            \wick{\c1{\bs{\nabla}{\cdots}\bs{\nabla}} \c1{\bs{S}}\bs{\nabla}{\cdots}\bs{\nabla}\bs{S}} &=\bs{\nabla}{\cdots}\bs{\nabla}\DD\bs{\nabla}{\cdots}\bs{\nabla} \bs{S}'\bs{\nabla}{\cdots}\bs{\nabla}\bs{S}=0\;,
            \\
            \wick{\bs{\nabla}{\cdots}\bs{\nabla} \c1{\bs{S}}\c1{\bs{\nabla}{\cdots}\bs{\nabla}}\bs{S}} &=\bs{\nabla}{\cdots}\bs{\nabla} \bs{S}'\bs{\nabla}{\cdots}\bs{\nabla}\DD\bs{\nabla}{\cdots}\bs{\nabla}\bs{S}=0\;,
           \\
            \wick{\bs{\nabla}{\cdots}\bs{\nabla} \c1{\bs{S}}\bs{\nabla}{\cdots}\bs{\nabla}\c1{\bs{S}}} &=\bs{\nabla}{\cdots}\bs{\nabla} \bs{S}'\bs{\nabla}{\cdots}\bs{\nabla}(\bs{S}\cdot\bs{l}^{\sharp})=0\;,
        \end{aligned}
        \end{equation}
        each of which vanish. Prime denotes the expression obtained after the removal of $\bs{l}$. The first case is zero due to vanishing trace of ${\bs{S}}$. In the remaining cases, we use ${\bs{\nabla l}=0}$, which allows us to move $\bs{l}$ anywhere in the expression; it gives rise either to the contraction ${\bs{S}\cdot\bs{l}^{\sharp}=0}$ (fourth line) or to derivative ${\DD=\bs{l}^{\sharp}\cdot\bs{\nabla}}$ (second and third lines). Thanks to \eqref{eq:DnablaR}, $\DD\bs{\nabla}{\cdots}\bs{\nabla}\bs{S}$ and $\DD\bs{\nabla}{\cdots}\bs{\nabla}\bs{S}'$ vanish.\footnote{It can be shown using \eqref{eq:NPspincoeffs} that $\DD\bs{S} = 0$ implies $\DD\bs{S}' = 0$ for CCNV $\bs{l}^\sharp$ in any frame (not necessarily the PP frame).}
        
    \item $\bs{\nabla}{\cdots}\bs{\nabla}\bs{S}\bs{\nabla}{\cdots}\bs{\nabla}\bs{C}$ vanishes if ${\bs{S}}$ has no free index. In addition to that, if at least one index of $\bs{S}$ is contracted with $\bs{C}$ then $\bs{C}$ must have at most one free index for the expression to vanish.

        As before, it is always ensured that at least one index of $\bs{S}$ corresponds to $\bs{l}$ (since $\bs{S}$ is of type III), which can be contracted back to $\bs{S}$, to derivatives, or to $\bs{C}$:
        \begin{equation}
        \begin{aligned}
            \bs{\nabla}{\cdots}\bs{\nabla}     \overset{\sqcap}{\bs{S}}\bs{\nabla}{\cdots}\bs{\nabla}\bs{C} &=0\;,
            \\
            \wick{\c1{\bs{\nabla}{\cdots}\bs{\nabla}} \c1{\bs{S}}\bs{\nabla}{\cdots}\bs{\nabla}\bs{C}} &=\bs{\nabla}{\cdots}\bs{\nabla}\DD\bs{\nabla}{\cdots}\bs{\nabla}\bs{S}'\bs{\nabla}{\cdots}\bs{\nabla}\bs{C}=0\;,
            \\
            \wick{\bs{\nabla}{\cdots}\bs{\nabla} \c1{\bs{S}}\c1{\bs{\nabla}{\cdots}\bs{\nabla}}\bs{C}} &=\bs{\nabla}{\cdots}\bs{\nabla} \bs{S}'\bs{\nabla}{\cdots}\bs{\nabla}\DD\bs{\nabla}{\cdots}\bs{\nabla}\bs{C}=0\;,
            \\
            \wick{\bs{\nabla}{\cdots}\bs{\nabla} \c1{\bs{S}}\bs{\nabla}{\cdots}\bs{\nabla}\c1{\bs{C}}} &=\bs{\nabla}{\cdots}\bs{\nabla} \bs{S}'\bs{\nabla}{\cdots}\bs{\nabla}(\bs{C}\cdot\bs{l}^{\sharp})\;.
        \end{aligned}
        \end{equation}
        The first three lines vanish for similar reasons as the first three lines of \eqref{eq:SSterms}, specifically because $\bs{S}$ is tracefree and ${\DD\bs{\nabla}{\cdots}\bs{\nabla}\bs{S}'=\DD\bs{\nabla}{\cdots}\bs{\nabla}\bs{C}=0}$ due to \eqref{eq:DnablaR}. However, this time the last line is non-zero in general since the contraction of $\bs{C}$ with $\bs{l}$ reads
        \begin{equation}\label{eq:C*l}
            \bs{C}\cdot\bs{l}^{\sharp} = \Psi_3 (\bs{l}\wedge\bs{m})\bs{l} +  \bar{\Psi}_3 (\bs{l}\wedge\bar{\bs{m}})\bs{l}\;.
        \end{equation}
        If this rank-3 tensor $\bs{C}\cdot\bs{l}^{\sharp}$ has at most one free index, then at least one of the remaining indices must be associated with $\bs{l}$. Let us inspect all possible types of contractions with this $\bs{l}$:
        \begin{equation}
        \begin{aligned}
            \bs{\nabla}{\cdots}\bs{\nabla} \bs{S}'\bs{\nabla}{\cdots}\bs{\nabla}    \overset{\sqcap}{(\bs{C}\cdot\smash{\bs{l}^{\sharp}})} &=0\;,
            \\
            \wick{\bs{\nabla}{\cdots}\bs{\nabla} \bs{S}'\c1{\bs{\nabla}{\cdots}\bs{\nabla}}\c1{(\bs{C}\cdot\smash{\bs{l}^{\sharp}})}} &=\bs{\nabla}{\cdots}\bs{\nabla} \bs{S}'\bs{\nabla}{\cdots}\bs{\nabla}\DD\bs{\nabla}{\cdots}\bs{\nabla}(\bs{C}\cdot\bs{l}^{\sharp})'=0\;,
            \\
            \wick{\c1{\bs{\nabla}{\cdots}\bs{\nabla}} \bs{S}'\bs{\nabla}{\cdots}\bs{\nabla}\c1{(\bs{C}\cdot\smash{\bs{l}^{\sharp}})}} &=\bs{\nabla}{\cdots}\bs{\nabla}\DD\bs{\nabla}{\cdots}\bs{\nabla} \bs{S}'\bs{\nabla}{\cdots}\bs{\nabla}(\bs{C}\cdot\bs{l}^{\sharp})'=0\;,
            \\
            \wick{\bs{\nabla}{\cdots}\bs{\nabla} \c1{\bs{S}'}\bs{\nabla}{\cdots}\bs{\nabla}\c1{(\bs{C}\cdot\smash{\bs{l}^{\sharp}})}} &= \bs{\nabla}{\cdots}\bs{\nabla} (\bs{S}'\cdot\bs{l}^{\sharp})\bs{\nabla}{\cdots}\bs{\nabla}(\bs{C}\cdot\bs{l}^{\sharp})'=0\;.
        \end{aligned}
        \end{equation}
        The first line vanishes since the trace of $\bs{C}\cdot\bs{l}^{\sharp}$ is zero. The contractions with derivatives (second and third lines) lead to vanishing expressions ${\DD\bs{\nabla}{\cdots}\bs{\nabla}(\bs{C}\cdot\bs{l}^{\sharp})' = \DD\bs{\nabla}{\cdots}\bs{\nabla}\bs{S}'=0}$. (Let us recall that $(\bs{C}\cdot\bs{l}^{\sharp})'$ is a rank-2 tensor that is obtained by stripping \eqref{eq:C*l} of $\bs{l}$.) Finaly, the last line involves ${\bs{S}'\cdot\bs{l}^{\sharp} = 0}$.

    \item $\bs{\nabla}{\cdots}\bs{\nabla}\bs{S}\bs{\nabla}{\cdots}\bs{\nabla}\bs{C}$ vanishes if ${\bs{C}}$ has no free index.
    
    From the decomposition of $\bs{C}$ in \eqref{eq:SCtypeIII}, we see that at least one $\bs{l}$ is contracted back to $\bs{C}$, to derivatives, or to~$\bs{S}$:
        \begin{equation}
        \begin{aligned}
            \bs{\nabla}{\cdots}\bs{\nabla}\bs{S} \bs{\nabla}{\cdots}\bs{\nabla}\overset{\sqcap}{\bs{C}} &=0\;,
            \\
            \wick{\c1{\bs{\nabla}{\cdots}\bs{\nabla}} \bs{S}\bs{\nabla}{\cdots}\bs{\nabla}\c1{\bs{C}}} &=\bs{\nabla}{\cdots}\bs{\nabla}\DD\bs{\nabla}{\cdots}\bs{\nabla}\bs{S}\bs{\nabla}{\cdots}\bs{\nabla}\bs{C}'=0\;,
            \\
            \wick{\bs{\nabla}{\cdots}\bs{\nabla} \bs{S}\c1{\bs{\nabla}{\cdots}\bs{\nabla}}\c1{\bs{C}}} &=\bs{\nabla}{\cdots}\bs{\nabla} \bs{S}\bs{\nabla}{\cdots}\bs{\nabla}\DD\bs{\nabla}{\cdots}\bs{\nabla}\bs{C}'=0\;,
            \\
            \wick{\bs{\nabla}{\cdots}\bs{\nabla} \c1{\bs{S}}\bs{\nabla}{\cdots}\bs{\nabla}\c1{\bs{C}}} &=\bs{\nabla}{\cdots}\bs{\nabla} (\bs{S}\cdot\bs{l}^{\sharp})\bs{\nabla}{\cdots}\bs{\nabla}\bs{C}'=0\;.
        \end{aligned}
        \end{equation}
        All the possibilities vanish since $\bs{C}$ is traceless, $\bs{S} \cdot \bs{l}^{\sharp} = 0$, and $\DD\bs{\nabla}{\cdots}\bs{\nabla}\bs{S}=\DD\bs{\nabla}{\cdots}\bs{\nabla}\bs{C}'=0$.

    \item $\bs{\nabla}{\cdots}\bs{\nabla}\bs{C}\bs{\nabla}{\cdots}\bs{\nabla}\bs{C}$ vanishes if one ${\bs{C}}$ has no free index. In addition to that, if at least one index of this $\bs{C}$ is contracted into the second $\bs{C}$ then the latter must have at most one free index for the expression to vanish.

        Without loss of generality, let us assume that $\bs{C}$ without free indices is the first one. As follows from \eqref{eq:SCtypeIII}, this (first) $\bs{C}$ has at least one $\bs{l}$ which can be contracted either back to $\bs{C}$, to derivatives, or to the other (second) $\bs{C}$:
        \begin{equation}
        \begin{aligned}
            \bs{\nabla}{\cdots}\bs{\nabla} \overset{\sqcap}{\bs{C}}\bs{\nabla}{\cdots}\bs{\nabla}\bs{C} &=0\;,
            \\
            \wick{\c1{\bs{\nabla}{\cdots}\bs{\nabla}} \c1{\bs{C}}\bs{\nabla}{\cdots}\bs{\nabla}\bs{C}} &=\bs{\nabla}{\cdots}\bs{\nabla}\DD\bs{\nabla}{\cdots}\bs{\nabla} \bs{C}'\bs{\nabla}{\cdots}\bs{\nabla}\bs{C}=0\;,
            \\
            \wick{\bs{\nabla}{\cdots}\bs{\nabla} \c1{\bs{C}}\c1{\bs{\nabla}{\cdots}\bs{\nabla}}\bs{C}} &=\bs{\nabla}{\cdots}\bs{\nabla} \bs{C}'\bs{\nabla}{\cdots}\bs{\nabla}\DD\bs{\nabla}{\cdots}\bs{\nabla}\bs{C}=0\;,
            \\
            \wick{\bs{\nabla}{\cdots}\bs{\nabla} \c1{\bs{C}}\bs{\nabla}{\cdots}\bs{\nabla}\c1{\bs{C}}} &=\bs{\nabla}{\cdots}\bs{\nabla} \bs{C}'\bs{\nabla}{\cdots}\bs{\nabla}(\bs{C}\cdot\bs{l}^{\sharp})\;.
        \end{aligned}
        \end{equation}
        The first three cases vanish due to the tracefreeness of $\bs{C}$ and \eqref{eq:DnablaR}. In the last case, the rank-3 tensor ${\bs{C} \cdot \bs{l}^{\sharp}}$ of the form \eqref{eq:C*l} has at most one free index and therefore at least one $\bs{l}$ is contracted either back to $\bs{C} \cdot \bs{l}^{\sharp}$, to derivatives, or to $\bs{C}'$:
        \begin{equation}
        \begin{aligned}
            \bs{\nabla}{\cdots}\bs{\nabla} \bs{C}'\bs{\nabla}{\cdots}\bs{\nabla}\overset{\sqcap}{(\bs{C}\cdot\smash{\bs{l}^{\sharp}})} &=0\;,
            \\
            \wick{\bs{\nabla}{\cdots}\bs{\nabla} \bs{C}'\c1{\bs{\nabla}{\cdots}\bs{\nabla}}\c1{(\bs{C}\cdot\smash{\bs{l}^{\sharp}})}} &=\bs{\nabla}{\cdots}\bs{\nabla} \bs{C}'\bs{\nabla}{\cdots}\bs{\nabla}\DD\bs{\nabla}{\cdots}\bs{\nabla}(\bs{C}\cdot\bs{l}^{\sharp})'=0\;,
            \\
            \wick{\c1{\bs{\nabla}{\cdots}\bs{\nabla}} \bs{C}'\bs{\nabla}{\cdots}\bs{\nabla}\c1{(\bs{C}\cdot\smash{\bs{l}^{\sharp}})}} &=\bs{\nabla}{\cdots}\bs{\nabla}\DD\bs{\nabla}{\cdots}\bs{\nabla} \bs{C}'\bs{\nabla}{\cdots}\bs{\nabla}(\bs{C}\cdot\bs{l}^{\sharp})'=0\;,
            \\
            \wick{\bs{\nabla}{\cdots}\bs{\nabla} \c1{\bs{C}\smash{{}'}}\bs{\nabla}{\cdots}\bs{\nabla}\c1{(\bs{C}\cdot\smash{\bs{l}^{\sharp}})}} &= \bs{\nabla}{\cdots}\bs{\nabla} (\bs{C}'\cdot\bs{l}^{\sharp})\bs{\nabla}{\cdots}\bs{\nabla}(\bs{C}\cdot\bs{l}^{\sharp})'\;.
        \end{aligned}
        \end{equation}
        The first three cases vanish for the same reasons as above. The rank-2 tensor ${\bs{C}'\cdot\bs{l}^{\sharp}}$ still has one $\bs{l}$ which is contracted either back to ${\bs{C}'\cdot\bs{l}^{\sharp}}$, to derivatives, or to $(\bs{C}\cdot\bs{l}^{\sharp})'$:
        \begin{equation}
        \begin{aligned}\label{eq:CCterms-last}
            \bs{\nabla}{\cdots}\bs{\nabla} \overset{\sqcap}{(\smash{\bs{C}'}\cdot\smash{\bs{l}^{\sharp}})}\bs{\nabla}{\cdots}\bs{\nabla}(\bs{C}\cdot\bs{l}^{\sharp})' &=0\;,
            \\
            \wick{\c1{\bs{\nabla}{\cdots}\bs{\nabla}} \c1{(\smash{\bs{C}'}\cdot\smash{\bs{l}^{\sharp}})}\bs{\nabla}{\cdots}\bs{\nabla}}(\bs{C}\cdot\bs{l}^{\sharp})' &=\bs{\nabla}{\cdots}\bs{\nabla} \DD\bs{\nabla}{\cdots}\bs{\nabla}(\bs{C}'\cdot\bs{l}^{\sharp})'\bs{\nabla}{\cdots}\bs{\nabla}(\bs{C}\cdot\bs{l}^{\sharp})'=0\;,
            \\
            \wick{\bs{\nabla}{\cdots}\bs{\nabla} \c1{(\smash{\bs{C}'}\cdot\smash{\bs{l}^{\sharp}})}\c1{\bs{\nabla}{\cdots}\bs{\nabla}}}(\bs{C}\cdot\bs{l}^{\sharp})' &=\bs{\nabla}{\cdots}\bs{\nabla} (\bs{C}'\cdot\bs{l}^{\sharp})'\bs{\nabla}{\cdots}\bs{\nabla}\DD\bs{\nabla}{\cdots}\bs{\nabla}(\bs{C}\cdot\bs{l}^{\sharp})'=0\;,
            \\
            \wick{\bs{\nabla}{\cdots}\bs{\nabla} \c1{(\smash{\bs{C}'}\cdot\smash{\bs{l}^{\sharp}})}\bs{\nabla}{\cdots}\bs{\nabla}\c1{(\bs{C}\cdot\smash{\bs{l}^{\sharp}})\smash{{}'}}} &= \bs{\nabla}{\cdots}\bs{\nabla} (\bs{C}'\cdot\bs{l}^{\sharp})'\bs{\nabla}{\cdots}\bs{\nabla}[(\bs{C}\cdot\bs{l}^{\sharp})'\cdot\bs{l}^{\sharp}]=0\;.
        \end{aligned}
        \end{equation}
        Again, the first three cases vanish as before and, in the last case, ${(\bs{C}\cdot\bs{l}^{\sharp})'\cdot\bs{l}^{\sharp} = 0}$ as follows for the Weyl tensor of type III from \eqref{eq:SCtypeIII}.
\end{itemize}

\subsection{Relevant scalars and rank-2 tensors}

Here, we focus on particular scalars and rank-2 tensors appearing in the field equations \eqref{eq:fieldeq}.

\begin{itemize}
    \item $\bs{\nabla^c \nabla^d} \square^n \bs{S_{cd}}$, $\bs{S^{cd}}\square^n \bs{S_{cd}}$, and ${\bs{C^{cdef}} \square^n \bs{C_{cdef}}}$ vanish.
    
        As mentioned above, type III pp-wave spacetimes are of VSI, meaning that all scalars constructed as contractions of $\bs{\nabla}{\cdots}\bs{\nabla}\bs{S}$ and $\bs{\nabla}{\cdots}\bs{\nabla}\bs{C}$ vanish. Remark that two latter scalars can also be shown to vanish using the results of Section \ref{ssec:TensorsQuadraticInCurvature}.
        
    \item $\bs{\nabla_c \nabla_{(a}} \square^n \bs{S_{b)}{}^c}$ and $\bs{S_{c(a}}\square^n \bs{S_{b)}{}^c}$ cancel each other.

        Commuting $\bs{\nabla^c}$ and $\bs{\nabla^d}$, it turns out that $\bs{\nabla_c \nabla_{(a}} \square^n \bs{S_{b)}{}^c}$ cancels exactly with $\bs{S_{c(a}}\square^n \bs{S_{b)}{}^c}$,
        \begin{equation}
            \bs{g_{d(a}} (\bs{\nabla^c \nabla^d} - \bs{S^{cd}}) \square^n \bs{S_{b)c}} = \bs{\nabla_{(a} \nabla^c} \square^n \bs{S_{b)c}} + \bs{S_{(a}{}^c} \square^n \bs{S_{b)c}} - \bs{C_{acbd}} \square^n \bs{S^{cd}} + \tfrac12 \bs{g_{ab} S_{cd}} \square^n \bs{S^{cd}} = 0\;,
        \end{equation}
        where we employed \eqref{eq:divboxS}, \eqref{eq:SboxS} and the fact that scalars constructed from $\bs{\nabla}{\cdots}\bs{\nabla}\bs{S}\bs{\nabla}{\cdots}\bs{\nabla}\bs{S}$ vanish.

    \item $\square^n \bs{S_{ab}}$ is of b.o.\ $-1$ because $\bs{\nabla}{\cdots}\bs{\nabla S}$ are balanced tensors.
    
    \item $(\bs{\nabla^c \nabla^d} + \tfrac12 \bs{S^{cd}}) \square^{n} \bs{C}_{\bs{c}\bs{(ab)}\bs{d}}$ takes the form \eqref{eq:nnSboxnC}.
    
        Recasting the contracted Bianchi identities with $R=0$ in terms of TF Ricci and Weyl, one obtains
        \begin{equation}\label{eq:contractedBianchi}
            \bs{\nabla_b S^b{}_a} = 0\;, \quad
            \bs{\nabla_{d}C_{abc}{}^{d}} = - \bs{\nabla_{[a}S_{b]c}}\;,
        \end{equation}
        and consequently, using \eqref{eq:SboxS},
        \begin{equation}\label{eq:div*contractedBianchi}
            \bs{\nabla^{b}\nabla^{d}C_{abcd}} =  \tfrac12 \left(\square \bs{S_{ac}} - \bs{S^{bd} C_{abcd}} \right)\;.
        \end{equation}    
        The term under consideration (i.e., the term involving $\mathcal{F}_3(\square)$ in \eqref{eq:fieldeq}) can be expressed recursively by commuting one $\square$ over $\bs{\nabla^c\nabla^d}$ as
        \begin{equation}\label{eq:F3termrecursive}
            (\bs{\nabla^c \nabla^d} + \tfrac12 \bs{S^{cd}})\square^{n+1} \bs{C_{c(ab)d}} =
            \square(\bs{\nabla^c\nabla^d} + \tfrac12 \bs{S^{cd}}) \square^{n} \bs{C_{c(ab)d}} - \bs{Q}_{n\bs{ab}}\;,
        \end{equation}
        with $\bs{Q}_{n}$ being rank-2 tensors of b.o.\ $-2$,
        \begin{equation}\label{eq:Qdef}
        \begin{aligned}
            \bs{Q}_{n\bs{ab}} &\equiv - \tfrac12\square(\bs{S_{(a}{}^c}\square^{n}\bs{S_{b)c}}) - \tfrac32 \bs{S_{(a}{}^{c}} \square^{n+1} \bs{S_{b)c}} + 3 \bs{\nabla_{c}S_{d(a} \nabla^d} \square^n \bs{S_{b)}{}^{c}} - \tfrac{9}{2} \bs{\nabla_d S_{c(a} \nabla^d} \square^n \bs{S_{b)}{}^c} \\
            &\feq - \square^n \bs{C_{acbd}} \square \bs{S^{cd}}
            - 4 \bs{\nabla^e\nabla^d}\square^n \bs{S_{(a}{}^c C_{b)dce}} - \bs{\nabla^e\nabla^d S_{(a}{}^c }\square^n \bs{C_{b)ecd}}
            \\
            &\feq -  2 \bs{\nabla_f}\square^n \bs{C_{dec(a} \nabla^c C_{b)}{}^{def}}
            - 2 \bs{\nabla_f} \square^n \bs{C_{dec(a} \nabla^e C_{b)}{}^{dcf}}\;.
        \end{aligned}
        \end{equation}
        To get this expression, we employed the contracted Bianchi identities \eqref{eq:contractedBianchi}, the results of Section \ref{ssec:TensorsQuadraticInCurvature}, equations \eqref{eq:divboxS}, \eqref{eq:NablacNabladBoxCabcd}, \eqref{eq:greenorangeterms}, \eqref{eq:cyanterms}, the fact that terms cubic in curvature vanish, and the fact that covariant derivatives commute (since the commutator introduces one more curvature tensor). Starting with \eqref{eq:div*contractedBianchi} and applying \eqref{eq:F3termrecursive} repeatedly, we finally obtain
        \begin{equation}\label{eq:nnSboxnC}
            (\bs{\nabla^c \nabla^d} + \tfrac12 \bs{S^{cd}}) \square^{n} \bs{\bs{C}_{\bs{c}(\bs{ab})\bs{d}}} = - \tfrac12 \square^{n+1} \bs{S_{ab}} -  \sum_{k=0}^{n-1} \square^k\bs{Q}_{(n-k-1)\bs{ab}}\;.
        \end{equation}
        Note that for ${n = 0}$ the sum is empty and we recover \eqref{eq:div*contractedBianchi}.
\end{itemize}


\section{Field equations for gyratons}\label{sec:feqgyr}

\subsection{Gyratons}

Let us now focus on specific geometries called \textit{gyratons}, which are known to describe various spinning null sources. Their subclass within type III pp-waves is given by the metric \cite{Podolsky:2014lpa}
\begin{equation}\label{eq:gyrmetric}
    \bs{g} = - \bs{\dd}u \vee \big(\bs{\dd}r + H\,\bs{\dd}u + J\,\bs{\dd}\varphi\big) + \bs{\dd}\rho\bs{\dd}\rho + \rho^2 \bs{\dd}\varphi\bs{\dd}\varphi\;,
\end{equation}
where $H=H(u,\rho, \varphi)$ and $J=J(u, \rho, \varphi)$ are two arbitrary functions that can be determined from the field equations. The coordinate $r$ is an affine parameter along the null congruence generated by CCNV ${\bs{\xi}=\bs{\pp}_r}$. The null coordinate $u$ is the retarded time for which ${\bs{\xi}^{\flat}=-\bs{\dd}u}$. The coordinates $\rho$ and $\varphi$ are polar coordinates spanning the 2-dimensional flat submanifolds of constant $u$ at each $r$. Occasionally, we will also use the Cartesian coordinates ${x=\rho\cos\varphi}$, ${y=\rho\sin\varphi}$, which are regular at the origin ${x=y=0}$ (i.e., $\rho=0$). The metric $\bs{g}$ then takes the form
\begin{equation}
    \bs{g}= -\bs{\dd}u \vee \big(\bs{\dd}r + H\,\bs{\dd}u -\tfrac{y}{\rho^2}J\bs{\dd}x+\tfrac{x}{\rho^2}J\bs{\dd}y\big) + \bs{\dd}x\bs{\dd}x +  \bs{\dd}y\bs{\dd}y\;.
\end{equation}

To utilize the NP formalism we need introduce the natural covector null frame,
\begin{equation}
    \bs{l}=-\bs{\dd}u\;,
    \quad
    \bs{n}=-\bs{\dd}r-H\bs{\dd}u-J\bs{\dd}\varphi\;,
    \quad
    \bs{m}=\frac{1}{\sqrt{2}}\left(\bs{\dd}\rho+i\rho\bs{\dd}\varphi\right)\;,
\end{equation}
and the corresponding dual vector frame,
\begin{equation}
    \bs{l}^{\sharp} = \bs{\partial}_r\;,
    \quad
    \bs{n}^{\sharp} = \bs{\partial}_u - H \bs{\partial}_r\;,
    \quad
    \bs{m}^{\sharp} = \frac{1}{\sqrt{2}} \bs{\pp}_{\rho} + \frac{i}{\sqrt{2}\rho}(\bs{\partial}_\varphi - J \bs{\partial}_r)\;.
\end{equation}
The vector $\bs{l}^{\sharp}$ is CCNV and the natural null frame is PP along geodesics generated by $\bs{l}^{\sharp}$. The spacetime has vanishing Ricci scalar, ${R=0}$. The non-vanishing components of the TF Ricci and Weyl tensors in this frame read
\begin{equation}\label{eq:Phi21Phi22Psi4gyr}
\begin{gathered}
    \Phi_{21} =-\Psi_3= -\frac{J_{,\rho\varphi}}{4\sqrt{2}\rho^2} + i \frac{J_{,\rho} - \rho J_{,\rho\rho}}{4\sqrt{2}\rho^2}\;,
    \quad
    \Phi_{22} = \frac{1}{2} \triangle H + \frac{\bigl(J_{,\rho}\bigr)^2 - 2 J_{,u\varphi}}{4 \rho^2}\;,
    \\
    \Psi_{4} =\frac{1}{2 \rho^2}\big( -\rho H_{,\rho}-2 i\rho H_{,\rho\varphi}+\rho^2 H_{,\rho\rho}+2 i H_{,\varphi}-H_{,\varphi\varphi}-2 i J_{,u}+J_{,u\varphi}+ i\rho J_{,u\rho}\big)\;,
\end{gathered}
\end{equation}
where we defined the Laplace operator on 2-dimensional transversal space,\footnote{Do not confuse with $\nDelta$-derivative of the NP formalism.}
\begin{equation}\label{eq:2dlapl}
    \triangle \equiv \pp_{\rho}^2+\frac{1}{\rho}\pp_{\rho} + \frac{1}{\rho^2} \partial^2_\varphi\;.
\end{equation}
The non-vanishing spin coefficients are
\begin{equation}
    \alpha = - \beta = - \frac{1}{2\sqrt{2} \rho}\;, 
    \quad 
    \mu = 2 \gamma = i \frac{J_{,\rho}}{2\rho}\;, 
    \quad 
    \nu = \frac{H_{,\rho}}{\sqrt{2}} - i \frac{H_{,\varphi} - J_{,u}}{\sqrt{2}\rho}\;.
\end{equation}
Note that $\lambda$ vanishes for gyratons \eqref{eq:gyrmetric} even though it is non-zero for general pp-waves of type III.

In what follows we will repeatedly use several properties of directional derivatives of the NP formalism. First, let us recall the properties of $\DD$-derivative. It annihilates all frame covectors \eqref{eq:Dnm} and curvature components \eqref{eq:CCNVBianchi}. Moreover, $\DD$ also commutes with all remaining derivatives \eqref{eq:Dcomppwave}. On the contrary, the commutators of $\delta$-derivatives (on scalars) are
\begin{equation}
    [\delta,\nDelta] =-\bar{\nu}\DD\;,
    \quad
    [\bar\delta,\delta] =-2\mu\DD-2\alpha\bar\delta+2\alpha\delta\;.
\end{equation}
The action of $\delta$ and $\bar{\delta}$ on the null frame is given by
\begin{equation}\label{eq:deltaframe}
\begin{aligned}
    \delta \bs{l} &= 0\;,
    &
    \delta \bs{m} &= - 2\alpha \bs{m}\;,
    \\
    \delta \bs{n} &= \mu \bs{m}\;,
    &
    \bar\delta \bs{m} &= -\mu \bs{l} +2\alpha \bs{m}\;.
\end{aligned}
\end{equation}
In addition, we will also need $\delta$-derivatives of the spin coefficients $\alpha$ and $\mu$,
\begin{equation}\label{eq:dalphadmu}
    \delta\alpha = \bar\delta\alpha = 2 \alpha^2\;, 
    \quad 
    \bar\delta\mu = -2 \Phi_{21}\;, 
    \quad 
    \delta\mu = 2 \bar{\Phi}_{21}\;.
\end{equation}
Finally, the action of $\nDelta$-derivative on the frame covectors is
\begin{equation}\label{eq:Deltaframe}
\begin{aligned}
    \nDelta \bs{l} &= 0\;,
    &
    \nDelta \bs{n} &=  \nu \bs{m} + \bar\nu \bar{\bs{m}}\;,
    &
    \nDelta \bs{m} &= \bar\nu \bs{l} + \mu \bs{m} \;.
\end{aligned}
\end{equation}

\subsection{Field equations}
The action of the wave operator $\square$ can be expressed in terms of the directional derivatives using the decomposition \eqref{eq:nablaDDeltadeltadelta} and the properties \eqref{eq:CCNVl}, \eqref{eq:Dnm}, \eqref{eq:deltaframe}, and \eqref{eq:Deltaframe}. We arrive at the formula
\begin{equation}\label{eq:sq}
    \square =-\nDelta\DD+\DD\nDelta+\bar{\delta}\delta+\delta\bar{\delta}-2\alpha\delta-2\alpha\bar{\delta}\;.
\end{equation}
For a scalar field $\phi$ subject to ${\DD \phi = 0}$, the wave operator reduces to 2-dimensional Laplace \eqref{eq:2dlapl},
\begin{equation}
    \square \phi = \triangle \phi\;,
\end{equation}
and $\delta$-derivative of $\phi$ is given by
\begin{equation}
    \delta \phi = \frac{\phi_{,\rho}}{\sqrt{2}} + i \frac{\phi_{,\varphi}}{\sqrt{2}\rho}\;.
\end{equation}
With the help of \eqref{eq:sq}, we can find the following useful formulas:
\begin{equation}\label{eq:boxphimmm}
\begin{aligned}
    \square(\phi\bs{m}) &=\big[\big(\triangle -4\alpha\bar{\delta}+4\alpha\delta -8\alpha^2\big)\phi\big]\bs{m}+\big[\big(-2\mu\delta -\delta\mu+4\alpha\mu \big)\phi\big]\bs{l}\;,
    \\
    \square(\phi \bar{\bs{m}}\bs{m} \bs{m}) &=\big[\big(\triangle -4\alpha\bar{\delta}+4\alpha\delta -8\alpha^2\big)\phi\big]\bar{\bs{m}}\bs{m}\bs{m}+\big[\big(-2\bar{\mu}\bar{\delta} -\bar{\delta}\bar{\mu}-4\alpha\bar{\mu} \big)\phi\big]\bs{l}\bs{m}\bs{m}
    \\
    &\feq +\big[\big(-2\mu\delta -\delta \mu+4\alpha \mu \big)\phi\big]\bar{\bs{m}}(\bs{l}\vee\bs{m})+[(2 \mu \bar{\mu}) \phi]\, \bs{l}(\bs{l}\vee\bs{m})\;.
\end{aligned}
\end{equation}
Applying the first one repeatedly on type III TF Ricci tensor \eqref{eq:SCtypeIII}, ${\square^n \bs{S} {=}-2\bs{l}\vee\square^n\big(\Phi_{21}\bs{m} {+}\bar{\Phi}_{21}\bar{\bs{m}}\big){+}2(\square^n\Phi_{22}) \bs{l}\bs{l}}$, we obtain
\begin{equation}
    \square^n \bs{S} 
    =-2\bs{l}\vee\big[\big(\mathsf{B}^n\Phi_{21}\big)\bs{m}+\big(\bar{\mathsf{B}}^n\bar{\Phi}_{21}\big)\bar{\bs{m}}\big]+2\bigg[\triangle^n\Phi_{22}+2\sum_{k=0}^{n-1}\triangle^k\left(\mathsf{M}\mathsf{B}^{n-k-1}\Phi_{21}+\bar{\mathsf{M}}\bar{\mathsf{B}}^{n-k-1}\bar{\Phi}_{21}\right)\bigg]\bs{l}\bs{l}\;,
\end{equation}
where we introduced the auxiliary differential operators
\begin{equation}
    \mathsf{B}\equiv \triangle -4\alpha\bar{\delta}+4\alpha\delta -8\alpha^2\;,
    \quad
    \mathsf{M}\equiv 2\mu\delta +\delta\mu-4\alpha\mu\;.
\end{equation}

In order to express \eqref{eq:nnSboxnC} explicitly, we also need to calculate tensors $\bs{Q}_{n}$. Using the above properties (particularly \eqref{eq:boxphimmm}), we can write the individual terms of \eqref{eq:Qdef} as
\begin{equation}
\begin{aligned}
    \bs{S_a{}^c} \square^{k+1} \bs{S_{bc}} &=  4 \big(\Phi_{21} \bar{\mathsf{B}}^{k+1} \bar{\Phi}_{21}+\bar{\Phi}_{21} \mathsf{B}^{k+1} \Phi_{21}\big)\bs{l_a l_b}\;,
    \\
    \bs{\nabla_{c}S_{da} \nabla^d} \square^k \bs{S_{b}{}^{c}} &= 4 \big[(\bar{\delta}\bar{\Phi}_{21} - 2 \alpha \bar{\Phi}_{21})(\bar{\delta}\bar{\mathsf{B}}^{k}\bar{\Phi}_{21} - 2 \alpha \bar{\mathsf{B}}^{k}\bar{\Phi}_{21})
    + (\bar{\delta}\Phi_{21} + 2 \alpha \Phi_{21}) (\delta \bar{\mathsf{B}}^{k}\bar{\Phi}_{21} + 2 \alpha \bar{\mathsf{B}}^{k}\bar{\Phi}_{21})
    \\
    &\feq + (\delta \Phi_{21} - 2 \alpha \Phi_{21})(\delta \mathsf{B}^{k}\Phi_{21} - 2 \alpha \mathsf{B}^{k}\Phi_{21})
    + (\delta \bar{\Phi}_{21} + 2 \alpha \bar{\Phi}_{21})(\bar{\delta}\mathsf{B}^{k} \Phi_{21} + 2 \alpha \mathsf{B}^{k}\Phi_{21})\big]\bs{l_a l_b}\;,
    \\
    \bs{\nabla_d S_{ca} \nabla^d} \square^k \bs{S_b{}^c} &= 4 \big[ ( \delta \Phi_{21}-2 \alpha \Phi_{21}) (\bar{\delta} \bar{\mathsf{B}}^k \bar\Phi_{21}-2 \alpha \bar{\mathsf{B}}^k \bar{\Phi}_{21})+ (\bar{\delta} \Phi_{21}+2 \alpha \Phi_{21})(\delta \bar{\mathsf{B}}^k \bar{\Phi}_{21}+2\alpha \bar{\mathsf{B}}^k \bar{\Phi}_{21})
    \\
    &\feq + ( \delta \bar{\Phi}_{21}+2 \alpha \bar{\Phi}_{21})(\bar{\delta} \mathsf{B}^k\Phi_{21}+2 \alpha \mathsf{B}^k \Phi_{21})+ ( \bar{\delta} \bar{\Phi}_{21}-2 \alpha \bar{\Phi}_{21})(\delta \mathsf{B}^k\Phi_{21}-2 \alpha \mathsf{B}^k \Phi_{21})\big]\bs{l_a l_b}\;,
    \\
    \square^k \bs{C_{acbd}} \square \bs{S^{cd}} &= 4 \big(\mathsf{B}^k\Phi_{21}\bar{\mathsf{B}}\bar{\Phi}_{21}+\bar{\mathsf{B}}^k\bar{\Phi}_{21}\mathsf{B}\Phi_{21}\big)\bs{l_a l_b}\;,
    \\
    \bs{\nabla^e\nabla^d} \square^k \bs{S_a{}^c C_{bdce}} &= - 2  \big[\Phi_{21}[(\delta^2 - 2 \alpha \delta - 4  \alpha^2) \mathsf{B}^k \Phi_{21} - (\delta \bar{\delta}- 2 \alpha \delta - 4 \alpha^2) \bar{\mathsf{B}}^k \bar{\Phi}_{21}]
    \\
    &\quad + \bar{\Phi}_{21}[( \bar{\delta}^2 - 2 \alpha  \bar{\delta} - 4 \alpha^2) \bar{\mathsf{B}}^k \bar\Phi_{21} - (\bar{\delta} \delta - 2 \alpha  \bar{\delta}- 4 \alpha^2) \mathsf{B}^k \Phi_{21}]\big]\bs{l_a l_b}\;,
    \\
    \bs{\nabla^e\nabla^d S_a{^c}} \square^k \bs{C_{becd}} &= - 2  \big[ \mathsf{B}^k \Phi_{21} (\delta \delta \Phi_{21}-2\alpha \delta \Phi_{21}-4 \alpha^2 \Phi_{21}-\delta \bar{\delta} \bar{\Phi}_{21}+2 \alpha \delta \bar{\Phi}_{21}+4 \alpha^2 \bar{\Phi}_{21})
    \\
    &\feq+\bar{\mathsf{B}}^k \bar{\Phi}_{21}(\bar{\delta} \bar{\delta} \bar{\Phi}_{21}-2 \alpha \bar{\delta} \bar{\Phi}_{21}-4 \alpha^2 \bar{\Phi}_{21}-\bar{\delta} \delta \Phi_{21}+2 \alpha \bar{\delta} \Phi_{21}+4 \alpha^2 \Phi_{21}) \big]\bs{l_a l_b}\;,
    \\
    \bs{\nabla^f} \square^k \bs{C_a{^{cde}} \nabla_c C_{bdef}} &=2 \big[ ( \delta \Phi_{21}-2 \alpha \Phi_{21}) (\delta \mathsf{B}^k \Phi_{21}-2 \alpha \mathsf{B}^k \Phi_{21})+ (\bar{\delta} \bar{\Phi}_{21}-2 \alpha \bar{\Phi}_{21})(\bar{\delta} \bar{\mathsf{B}}^k \bar{\Phi}_{21}-2\alpha \bar{\mathsf{B}}^k \bar{\Phi}_{21})\big]\bs{l_a l_b}\;,
    \\
    \bs{\nabla^f} \square^k \bs{C_a{}^{cde} \nabla_e C_{bdcf}} &=  \big[(\bar{\delta}\Phi_{21} + 2 \alpha \Phi_{21})(\delta \bar{\mathsf{B}}^{k}{}\bar{\Phi}_{21} + 2 \alpha \bar{\mathsf{B}}^{k}\bar{\Phi}_{21}) + (\delta \bar{\Phi}_{21} + 2 \alpha \bar{\Phi}_{21})(\bar{\delta}\mathsf{B}^{k}\Phi_{21} + 2 \alpha \mathsf{B}^{k}\Phi_{21})
    \\
    &\feq +(\delta \Phi_{21} - 2 \alpha \Phi_{21} + 2 \bar{\delta}\bar{\Phi}_{21} - 4 \alpha \bar{\Phi}_{21})(\bar{\delta}\bar{\mathsf{B}}^{k}\bar{\Phi}_{21} - 2 \alpha \bar{\mathsf{B}}^{k}\bar{\Phi}_{21})
    \\
    &\feq +(\bar{\delta}\bar{\Phi}_{21} - 2 \alpha \bar{\Phi}_{21} + 2 \delta \Phi_{21} - 4 \alpha \Phi_{21})(\delta \mathsf{B}^{k}\Phi_{21} - 2 \alpha \mathsf{B}^{k}\Phi_{21})\big]\bs{l_a l_b}\;.
\end{aligned}
\end{equation}
After adding all these terms together, we arrive at the compact expression
\begin{equation}
    \bs{Q}_{k} =\big(\mathsf{N}\mathsf{B}^{k}\Phi_{21}+\bar{\mathsf{N}}\bar{\mathsf{B}}^{k}\bar{\Phi}_{21}\big)\bs{ll}\;,
\end{equation}
where we introduced another operator
\begin{equation}
\begin{aligned}
    \mathsf{N} &= -2\big[ 4\bar{\Phi}_{21} \mathsf{B} -4 \Phi_{21} \delta^2  + 4 \bar{\Phi}_{21} \bar{\delta} \delta
    - 2(5 \delta \Phi_{21} - 14 \alpha \Phi_{21} - 5 \bar{\delta} \bar{\Phi}_{21} + 10 \alpha \bar{\Phi}_{21}) \delta
    + 4 \delta \bar{\Phi}_{21} \bar{\delta}
    \\
    &\feq +3\bar{\mathsf{B}}\bar{\Phi}_{21} - \delta^2 \Phi_{21} + \delta \bar{\delta} \bar{\Phi}_{21} + 22 \alpha \delta \Phi_{21} - 20 \alpha^2 \Phi_{21} - 20 \alpha \bar{\delta} \bar{\Phi}_{21} + 36 \alpha^2 \bar{\Phi}_{21} + 6 \alpha \delta \bar{\Phi}_{21}\big]\;.
\end{aligned}
\end{equation}

The field equations can be satisfied only if the cosmological constant $\Lambda$ vanish (because ${R=0}$) and the energy-momentum tensor $\bs{T}$ is of the algebraic type III,
\begin{equation}\label{eq:Tgyraton}
\begin{aligned}
    \bs{T} &= -2 \bs{l}\vee\big[\Xi_{21}\bs{m}+ \bar{\Xi}_{21}\bar{\bs{m}}\big]+ 2\Xi_{22}\bs{l}\bs{l}
    \\
    &=2\sqrt{2}\,\bs{\dd}u\vee\big[\Re{\Xi_{21}}\,\bs{\dd}\rho-\Im{\Xi_{21}}\rho\,\bs{\dd}\varphi\big]+2\Xi_{22}\,\bs{\dd}u\bs{\dd}u
    \\
    &=2\sqrt{2}\rho^{-1}\,\bs{\dd}u\vee\left[\big(x\Re{\Xi_{21}}+y\Im{\Xi_{21}}\big)\bs{\dd}x+\big(y\Re{\Xi_{21}}-x\Im{\Xi_{21}}\big)\bs{\dd}y\right]+2\Xi_{22}\,\bs{\dd}u\bs{\dd}u\;,
\end{aligned}
\end{equation}
where we introduced the components $\Xi_{21}$ and $\Xi_{22}$ in analogy to the notation for components of TF Ricci tensor, cf. \eqref{eq:SCtypeIII}. The resulting field equations for the gyraton metric take the form
\begin{equation}\label{eq:fieldeqgyr}
\boxed{
\begin{aligned}
    \big[1+\varkappa\mathcal{F}_2(\mathsf{B})\mathsf{B}\big]\Phi_{21} &=\varkappa \Xi_{21}\;,
    \\
    \big[1+\varkappa\left(\mathcal{F}_2(\triangle)+2\mathcal{F}_3(\triangle)\right)\triangle\big]\Phi_{22}+2\varkappa\sum_{n=1}^{\infty}\sum_{k=0}^{n-1}\triangle^k\big[\left(f_{2,n-1}\mathsf{M}+f_{3,n}\mathsf{N}\right)\mathsf{B}^{n-k-1}\Phi_{21}\phantom{\big]}
    \\
    +\left(f_{2,n-1}\bar{\mathsf{M}}+f_{3,n}\bar{\mathsf{N}}\right)\bar{\mathsf{B}}^{n-k-1}\bar{\Phi}_{21}\big] &=\varkappa \Xi_{22}\;.
\end{aligned}
}
\end{equation}
Upon inserting the components of the TF Ricci $\Phi_{21}$ and $\Phi_{22}$ from \eqref{eq:Phi21Phi22Psi4gyr} in \eqref{eq:fieldeqgyr}, we see that the two field equations are partly decoupled. Indeed, the first equation of \eqref{eq:fieldeqgyr} is independent of $H$, so we can find $J$ from this equation. With the obtained $J$, we can then calculate the corresponding operators $\mathsf{M}$ and $\mathsf{N}$ that appear in the second equation of \eqref{eq:fieldeqgyr} and solve it for $H$. Due to the linearity of the first equation in $J$ and the second equation in $H$, we may rely on the theorems for the existence of the solutions and make use of known mathematical methods for linear partial differential equations.

\subsection{Axial symmetry}\label{ssec:axialsym}
The field equations reduce even further if we assume the axial symmetry described by the Killing vector~${\bs{\pp}_{\varphi}}$,
\begin{equation}
    \pounds_{\bs{\pp}_{\varphi}}\bs{g}=0\;.
\end{equation}
This will not only make the functions $H$ and $J$ independent of $\varphi$, but also the derivatives $\pp_u$ will drop out from the field equations. As a result, we will be left with ordinary differential equations in coordinate $\rho$ (with additional trivial dependence on $u$). To arrive at this result, we first notice that the axial symmetry significantly simplifies the operators defined above (note that ${\delta=\bar{\delta}}$),
\begin{equation}\label{eq:opsimplif}
    \triangle=\triangle_0\;, 
    \quad
    \mathsf{B}=\bar{\mathsf{B}}=\triangle_1\;,
    \quad
    \mathsf{M}=-\bar{\mathsf{M}}=\diamondsuit/4\;,
    \quad
    \mathsf{N}=-\bar{\mathsf{N}}=\heartsuit/4\;,
\end{equation}
and the components of the TF Ricci tensor,
\begin{equation}\label{eq:Phiaxisym}
\begin{aligned}
    \Phi_{21} &=-\bar{\Phi}_{21} = \circledcirc J\;,
    \\
    \Phi_{22} &=\frac12\triangle_0 H+\frac{\big(J_{,\rho}\big)^2}{4 \rho^2}\;.
\end{aligned}
\end{equation}
In these equations, we introduced $\triangle_w$, which denotes the \textit{Bessel operators of order $w$}, and three other auxiliary ordinary differential operators ${\circledcirc}$, $\diamondsuit$, and $\heartsuit$,
\begin{equation}\label{eq:operators}
\begin{aligned}
    \triangle_w &\equiv \pp_{\rho}^2+\frac{1}{\rho}\pp_{\rho}-\frac{w^2}{\rho^2}\;,
    &
    \diamondsuit  &\equiv\frac{4i}{\sqrt{2}}\left(\frac{J_{,\rho}}{\rho}\pp_{\rho}+\frac{ J_{,\rho\rho}}{2\rho}+\frac{J_{,\rho}}{2\rho^2}\right)\;,
    \\
    {\circledcirc} &\equiv-\frac{i}{4\sqrt{2}}\pp_{\rho}\left(\frac{1}{\rho}\pp_{\rho}\right)\;, 
    & 
    \heartsuit &\equiv - 4i \sqrt{2} \left( 2\frac{\rho J_{,\rho\rho} - J_{,\rho}}{\rho^2} \partial^2_{\rho} + \frac{3 \rho^2 J_{,\rho\rho\rho} - 2 \rho J_{,\rho\rho} + 2 J_{,\rho}}{\rho^3} \partial_\rho + \frac{J_{,\rho\rho\rho\rho}}{\rho} \right)\;.
    \end{aligned}
\end{equation}
Furthermore, it follows from the first equation of \eqref{eq:fieldeqgyr} that the components of the energy-momentum tensor $\bs{T}$ must also obey ${\Xi_{21} =-\bar{\Xi}_{21}}$ to match the left-hand side; therefore, $\bs{T}$ takes the form
\begin{equation}\label{eq:Tgyraton2}
\begin{aligned}
    \bs{T} &= -2 \Xi_{21}\,\bs{l}\vee\left(\bs{m}-\bar{\bs{m}}\right)+ 2\Xi_{22}\,\bs{l}\bs{l}
    \\
    &=2\sqrt{2}i \,\Xi_{21}\rho\,\bs{\dd}u\vee\bs{\dd}\varphi+2\Xi_{22}\,\bs{\dd}u\bs{\dd}u
    \\
    &=2\sqrt{2}i\,\Xi_{21}\rho^{-1}\bs{\dd}u\vee\left(-y\bs{\dd}x+x\bs{\dd}y\right)+2\Xi_{22}\,\bs{\dd}u\bs{\dd}u\;.
\end{aligned}
\end{equation}
Using \eqref{eq:opsimplif} and \eqref{eq:Tgyraton2}, the field equations \eqref{eq:fieldeqgyr} finally reduce to
\begin{equation}\label{eq:fieldeqaxisym}
\boxed{
\begin{aligned}
    \big[1+\varkappa\mathcal{F}_2(\triangle_1)\triangle_1\big]\Phi_{21} &=\varkappa \Xi_{21}\;,
    \\
    \big[1+\varkappa\left(\mathcal{F}_2(\triangle_0)+2\mathcal{F}_3(\triangle_0)\right)\triangle_0\big]\Phi_{22}+\varkappa\sum_{n=1}^{\infty}\sum_{k=0}^{n-1}\triangle_0^k(f_{2,n-1}\diamondsuit+f_{3,n}\heartsuit)\triangle_1^{n-k-1}\Phi_{21}&=\varkappa \Xi_{22}\;.
\end{aligned}}
\end{equation}
These equations along with \eqref{eq:Phiaxisym} form a set of field equations for axially symmetric gyratons that we will study in the rest of the paper.

Notice that \eqref{eq:fieldeqaxisym} are ordinary differential equations for $H$ and $J$ in variable $\rho$ with additional (non-derivative) dependence on $u$. The partial linearity and decoupling of the equations enables us to split the general solution in two parts:
\begin{equation}
\begin{aligned}
    J &=J_{\textrm{hom}}+J_{\textrm{part}}[\Xi_{21}]\;,
    \\
    H &=H_{\textrm{hom}}+H_{\textrm{part}}[J,\Xi_{22}]\;.
\end{aligned}
\end{equation}
The \textit{homogeneous parts} $J_{\textrm{hom}}$ and $H_{\textrm{hom}}$ stand for all solutions of the homogeneous equations obtained by keeping the linear terms (in $J$ of the first equation and $H$ of the second equation),
\begin{equation}\label{eq:homparteq}
\begin{aligned}
    \big[1+\varkappa\mathcal{F}_2(\triangle_1)\triangle_1\big]{\circledcirc} J_{\textrm{hom}} &=0\;,
    \\
    \big[1+\varkappa\left(\mathcal{F}_2(\triangle_0)+2\mathcal{F}_3(\triangle_0)\right)\triangle_0\big]\triangle_0 H_{\textrm{hom}} &=0\;.
\end{aligned}
\end{equation}
The \textit{particular parts} $J_{\textrm{part}}$ and $H_{\textrm{part}}$ denote solutions satisfying the full inhomogeneous equations. It is important to emphasize that the homogeneous part $H_{\textrm{hom}}$ does not necessarily correspond to the solutions in the vacuum because $H_{\textrm{part}}$ may actually be non-trivial (different from any $H_{\textrm{hom}}$) even for ${\bs{T}=0}$. This is because the equation for $H_{\textrm{part}}$ also depends on $J$ obtained from the first equation. Through this dependence it is affected not only by $J_{\textrm{part}}$ (i.e., by $\Xi_{21}$), but also by $J_{\textrm{hom}}$.

Since the Bessel operators ${\triangle}_w$ arise naturally for the axially symmetric source, it turns out to be practical to introduce the \textit{Hankel transform of the order $w$} (see e.g. \cite{poularikas2000the}),\footnote{Hankel transforms of various functions can be found in tables \cite{Bateman_1954,gradshteyn2007}.}
\begin{equation}
    \mathcal{H}_w[\phi](s)=\int_{\mathbb{R}^+} \!\! d\rho\, \rho\, \phi(\rho)J_w(\rho s)\;,
    \quad
    \mathcal{H}_w^{-1}[\psi](\rho)=\int_{\mathbb{R}^+} \!\! d s\, s\, \psi(s)J_w(s\rho)\;.
\end{equation}
The reason is because the Bessel functions $J_w(\rho s)$ are the eigenfunctions of ${\triangle}_w$ with the eigenvalues $-s^2$; as a consequence of which the Hankel transforms of ${\triangle}_w$ are simply
\begin{equation}
    \mathcal{H}_w[\triangle_w\phi](s) =-s^2\mathcal{H}_w[\phi](s)\;.
\end{equation}

One method that can be used to find particular parts $J_{\textrm{part}}$ and $H_{\textrm{part}}$ is to first solve the first equation of \eqref{eq:fieldeqaxisym} for $\Phi_{21}^{\textrm{part}}$ using the Hankel transform of the order 1,
\begin{equation}\label{eq:Phi21h}
    \Phi_{21}^{\textrm{part}}=\varkappa\mathcal{H}_1^{-1}\bigg[\frac{\mathcal{H}_1[\Xi_{21}](\tilde{s})}{1-\varkappa\mathcal{F}_2(-\tilde{s}^2)\tilde{s}^2}\bigg](\rho)\;.
\end{equation}
Then we can obtain $J_{\textrm{part}}$ by integrating the first equation in \eqref{eq:Phiaxisym}, i.e.,
\begin{equation}
    {\circledcirc} J_{\textrm{part}}=\Phi_{21}^{\textrm{part}}\;.
\end{equation}
Its solution can be written in the form
\begin{equation}\label{eq:partJcf}
    J_{\textrm{part}}=2\sqrt{2}i\int_0^{\rho}\!\!d\rho'\,(\rho^2-\rho'^2)\Phi_{21}^{\textrm{part}}(\rho')\;,
\end{equation}
where we used the Cauchy formula for repeated integration. After choosing one specific function ${J=J_{\textrm{hom}}+J_{\textrm{part}}}$ (with desired asymptotic behavior, etc.), we get the explicit form of the operators $\diamondsuit$ and $\heartsuit$ from \eqref{eq:operators}.

A particular part $H_{\textrm{part}}$ is then obtained by solving the second equation of \eqref{eq:fieldeqaxisym}. This can be done again in two steps. First, we apply the Hankel transforms of orders 0 and 1 to get
\begin{equation}\label{eq:Phi22h}
\begin{aligned}
    \Phi_{22}^{\textrm{part}} &= \varkappa\mathcal{H}_0^{-1}\left[\frac{\mathcal{H}_0[\Xi_{22}](s)}{1-\varkappa\left(\mathcal{F}_2(-s^2)+2\mathcal{F}_3(-s^2)\right)s^2} \right](\rho)
    \\
    &\feq -\varkappa\sum_{n=1}^{\infty}\sum_{k=0}^{n-1}\mathcal{H}_0^{-1}\left[\frac{(-s^2)^k\mathcal{H}_0\left[\left(f_{2,n-1}\diamondsuit+f_{3,n}\heartsuit\right)\mathcal{H}_1^{-1}\left[(-\tilde{s}^2)^{n-k-1}\mathcal{H}_1[\Phi_{21}](\tilde{s})\right](\rho)\right](s)}{1-\varkappa\left(\mathcal{F}_2(-s^2)+2\mathcal{F}_3(-s^2)\right)s^2}\right](\rho)\;.
\end{aligned}
\end{equation}
Then we employing \eqref{eq:Phiaxisym} and arrive at the equation for $H_{\textrm{part}}$,
\begin{equation}\label{eq:Poisson}
    \triangle_0 H_{\textrm{part}}=2\Phi_{22}^{\textrm{part}} -\frac{\big(J_{,\rho}\big)^2}{2 \rho^2}\equiv W(u,\rho)\;,
\end{equation}
in which we recognize Poisson's equation with an axially symmetric right-hand side $W(u,\rho)$. It can be solved using convolution with the Green's function (integrated out over angles $\varphi$),
\begin{equation}\label{eq:partHgf}
\begin{aligned}
    H_{\textrm{part}}=G\star W &=\frac{1}{4\pi}\int_{0}^\infty\!\!\int_{0}^{2\pi}\!\!d\rho'd\varphi'\rho'\log\left(\frac{\rho ^2+\rho'^2-2 \rho  \rho' \cos (\varphi -\varphi')}{\rho_0^2}\right) W(u,\rho')
    \\
    &=\int_{0}^\infty\!\!d\rho'\,\rho'L(\rho,\rho')W(u,\rho')\;,
\end{aligned}
\end{equation}
where
\begin{equation}
    L(\rho,\rho')\equiv\begin{cases}
        \log\big(\rho'/\rho_0\big)\;, &\rho<\rho'\;,
        \\
        \log\big(\rho/\rho_0\big)\;, &\rho>\rho'\;.
    \end{cases}
\end{equation}
Finally, let us mention that these generic methods assume convergence of certain integrals. If these assumptions are not satisfied, one has to use different techniques as we will also need to do in one example bellow.


\section{Gyratons in GR}\label{sec:gyrgr}

Before we move on to the application in higher derivative gravity theories, we focus on the general relativity. We will review a known vacuum solution (in the notation of this paper) and also discuss a non-vacuum solution obtained by regularization of its Dirac-delta source. The Einstein--Hilbert action corresponds to setting all form-factors to zero,
\begin{equation}
    \mathcal{F}_1(\square)=\mathcal{F}_2(\square)=\mathcal{F}_3(\square)=0\;.
\end{equation}
Then the field equations for axially symmetric gyratons read 
\begin{equation}\label{eq:feqGR}
\begin{aligned}
    \Phi_{21}=\varkappa\Xi_{21}\;,
    \\
    \Phi_{22}=\varkappa\Xi_{22}\;,
\end{aligned}
\end{equation}
where $\Phi_{21}$ and $\Phi_{22}$ should be understood in terms of $J$ and $H$ through \eqref{eq:Phiaxisym}.

\subsection{Homogeneous parts}
It is instructive to first focus on the homogeneous parts $J_{\textrm{hom}}$ and $H_{\textrm{hom}}$. Following \eqref{eq:homparteq}, these functions satisfy two independent second order differential equations
\begin{equation}\label{eq:JHhomeq}
\begin{aligned}
    {\circledcirc} J_{\textrm{hom}} &=0\;,
    \\
    \triangle_0 H_{\textrm{hom}} &=0\;.
\end{aligned}
\end{equation}
which can be easily integrated out,
\begin{equation}\label{eq:JHhomgr}
\begin{aligned}
    J_{\textrm{hom}} &=c_1(u)\rho^2+c_2(u)\;,
    \\
    H_{\textrm{hom}} &=c_3(u)\log\rho +c_4(u)\;,
\end{aligned}
\end{equation}
where $c_i(u)$ denote four arbitrary functions of the null coordinate $u$. To clarify the meaning of \eqref{eq:JHhomgr}, we have to treat $J_{\textrm{hom}}$ and $H_{\textrm{hom}}$ in the language of distributions. For this purpose, we switch to Cartesian coordinates, which are well defined at the origin ${\rho=0}$ (unlike the polar coordinates). In these coordinates, the operators $\triangle_0$ and ${\circledcirc}$ from \eqref{eq:operators} are given by formulas
\begin{equation}
\begin{aligned}
    \triangle_0 H &=\left(\pp_x^2+\pp_y^2\right)H\;,
    \\
    {\circledcirc} J &=-\frac{i}{4\sqrt{2}}\big(\tfrac{x}{\rho}\pp_x+\tfrac{y}{\rho}\pp_y\big)\big[\pp_x\big(\tfrac{x}{\rho^2}J\big)+\pp_y\big(\tfrac{y}{\rho^2}J\big)\big]\;.
\end{aligned}
\end{equation}
Taking into consideration the distributional identity
\begin{equation}\label{eq:distridentity}
    \triangle_0 \log\rho =
    \pp_x\big(\tfrac{x}{\rho^2}\big)+\pp_y\big(\tfrac{y}{\rho^2}\big) =2\pi\delta(x)\delta(y)\;,
\end{equation}
we can now evaluate the action of the operators $\triangle_0$ and ${\circledcirc}$ on the homogeneous parts \eqref{eq:JHhomgr},
\begin{equation}\label{eq:homdistrgr}
\begin{aligned}
    {\circledcirc} J_{\textrm{hom}} &=-\frac{i\pi c_2(u)}{2\sqrt{2}}\bigg[\frac{y}{\rho}\delta(x)\delta'(y)+\frac{x}{\rho}\delta'(x)\delta(y)\bigg]\;,
    \\
    \triangle_0 H_{\textrm{hom}} &= 2\pi c_3(u)\delta(x)\delta(y)\;.
\end{aligned}
\end{equation}
Thus, we see that the functions ${J_{\textrm{hom}}=c_2(u)}$ and ${H_{\textrm{hom}}=c_3(u)\log\rho}$ may be considered as homogeneous parts only for ${\rho>0}$ (when treated as functions), but not in the distributional sense. With this in mind, we can now proceed to solutions for various sources.

\subsection{Vacuum}

Let us first look for solutions of \eqref{eq:feqGR} in the region with no matter content, ${\bs{T}=0}$. The condition ${\Xi_{21}=0}$ implies $J=J_{\hom}$, which, after taking into account ${\Xi_{22}=0}$, leads to a general solution
\begin{equation}
\begin{aligned}
    J &=c_1(u)\rho^2+c_2(u)\;,
    \\
    H &=c_3(u)\log\rho +c_4(u)-\tfrac{1}{2}c_1(u)^2\rho^2\;.
\end{aligned}
\end{equation}
The function $c_1(u)$ can be removed by a coordinate transformation ${\tilde{\varphi}=\varphi-\int\!\dd u\,c_1(u)}$ and the function $c_4(u)$ is of no physical relevance \cite{Griffiths:2009dfa}. Consequently, the solution can be equivalently rewritten in the form \cite{Bonnor:1970sb,Frolov:2005in}
\begin{equation}\label{eq:vacuumsolgr}
\begin{aligned}
    J &=\frac{\varkappa\chi_{\textrm{J}}(u)}{4\pi}\;,
    \\
    H &=\frac{\varkappa\chi_{\textrm{H}}(u)}{4\pi}\log\left(\frac{\rho^2}{\rho_0^2}\right)\;,
\end{aligned}
\end{equation}
where $\chi_{\textrm{J}}(u)$ and $\chi_{\textrm{H}}(u)$ are the \textit{profile functions} and $\rho_0$ is an arbitrary constant (without any physical significance). 

In order to interpret this solution, we will calculate the energy-momentum tensor $\bs{T}$ in a distributional sense. If we use the distributional formulas \eqref{eq:homdistrgr} together with ${J_{,\rho}/\rho=\pp_x(x J/\rho^2)+\pp_y(y J/\rho^2)}$ and \eqref{eq:distridentity}, we obtain
\begin{equation}\label{eq:distrsource}
\begin{aligned}
    \Xi_{21} &=-\frac{i\chi_{\textrm{J}}(u)}{2^3\sqrt{2}}\bigg[\frac{y}{\rho}\delta(x)\delta'(y)+\frac{x}{\rho}\delta'(x)\delta(y)\bigg]\;,
    \\
    \Xi_{22} &=\frac{\chi_{\textrm{H}}(u)}{2}\delta(x)\delta(y)+\frac{\varkappa\chi_{\textrm{J}}^2(u)}{2^4}(\delta(x)\delta(y))^2\;.
\end{aligned}
\end{equation}
The ill-defined term $(\delta(x)\delta(y))^2$ obviously arises because of the naive application of linear distributions to non-linear expressions \cite{Frolov:2005in}. It signifies that the distributional Dirac-delta sources describing null particles can be used only in the non-spinning case, ${\chi_{\textrm{J}}=0}$, or in the \textit{linearized regime of slow rotation}, ${O(\chi_{\textrm{J}}^2)\approx 0}$. In other words, the spinning null sources in the full theory must be spatially distributed. One possibility to overcome this issue is to glue the exterior vacuum solution \eqref{eq:vacuumsolgr} to an interior non-vacuum solution representing the spinning cylindrical beam of light of finite radius~\cite{Bonnor:1970sb}. Another option, employed here, is to consider Gaussian-type distribution of the spinning null matter that is non-zero throughout the spacetime. 

\subsection{Gaussian beam}

Motivated by \eqref{eq:distrsource}, we may obtain the Gaussian-type source by regularizing Dirac-delta distribution $\delta(x)$ using the \textit{nascent delta function} $\delta_{\epsilon}(x)$ given by the heat kernel, i.e., the Gaussian function,
\begin{equation}
    \delta_{\epsilon}(x)=\frac{e^{-x^2/4\epsilon^2}}{2\sqrt{\pi}\epsilon}\;.
\end{equation}
Here, the parameter ${\epsilon>0}$ controls the \textit{width} of the Gaussian. Replacing $\delta(x)$ by $\delta_{\epsilon}(x)$ in \eqref{eq:distrsource}, we get the energy-momentum tensor
\begin{equation}\label{eq:gausssource}
\begin{aligned}
    \Xi_{21} &=\frac{i\chi_{\textrm{J}}(u)}{2^6 \sqrt{2} \pi  \epsilon^4}\rho e^{-\rho ^2/4\epsilon^2}\;,
    \\
    \Xi_{22} &= \frac{\chi_{\textrm{H}}(u)}{2^3 \pi  \epsilon^2 } e^{-\rho ^2/4\epsilon^2}+\frac{\varkappa  \chi_{\textrm{J}}^2(u) }{2^8 \pi ^2 \epsilon ^4}e^{-\rho ^2/2\epsilon^2}\;.
\end{aligned}
\end{equation}
This specific choice of regularization will prove very useful in evaluating the Hankel transforms of the relevant functions that would otherwise be very difficult if not impossible. The corresponding function $J$ can be obtained from the formula for particular part \eqref{eq:partJcf},
\begin{equation}\label{eq:JgausGR}
    J=\frac{\varkappa  \chi_{\textrm{J}}(u)}{4 \pi }\Big(1-e^{-\frac{\rho ^2}{4 \epsilon ^2}}\Big)\;,
\end{equation}
where we subtracted the homogeneous part proportional to $\rho^2$ to match the asymptotic behavior at $\rho\to\infty$ with the vacuum solution \eqref{eq:vacuumsolgr}.

Having found $J$, we can now calculate $H$ by means of Green's function \eqref{eq:partHgf} and using the identity
\begin{equation}\label{eq:convgauss}
    G\star e^{-b\rho^2}=\frac{1}{4 b}\left[\log \left(\rho^2/\rho_0^2\right)-\Ei\left(-b \rho ^2\right)\right]\;,
\end{equation}
with $b$ being a positive constant. The result is
\begin{equation}\label{eq:HgausGR}
    H =\frac{\varkappa\chi_{\textrm{H}}(u)}{4 \pi }  \left[\log \left(\frac{\rho ^2}{\rho_0^2}\right)-   \Ei\left(-\frac{\rho ^2}{4 \epsilon ^2}\right)\right]\;.
\end{equation}
Notice that the regularization of the source induces the behavior ${J=O(\rho^2)}$ and ${H=O(1)}$ near ${\rho=0}$. Finally, let us point out that the parameter ${\epsilon}$, characterizing the width of the Gaussian beam, should not be regarded as infinitesimal but as a finite quantity. In fact, due to the presence of $(\delta_{\epsilon}(x)\delta_{\epsilon}(y))^2$ term, many expressions are expected to blow up in the limit ${\epsilon\to0}$ unless we neglect the non-linear terms proportional to ${O(\chi_{\mathrm{J}}^2)}$.


\section{Gyratons in SG}\label{sec:gyrsg}

The Stelle gravity is obtained if we set the form-factors to constants \cite{Stelle:1977ry,Deser:2011xc},
\begin{equation}\label{eq:Fconst}
    \mathcal{F}_{1}(\square)= \alpha+\beta/4\;,
    \quad
    \mathcal{F}_{2}(\square)=\beta\;,
    \quad
    \mathcal{F}_{3}(\square)=0\;.
\end{equation}
where, without loss of generality, we left out the Weyl term. It is always possible to achieve $\mathcal{F}_{3}(\square)=0$ by adding the Gauss--Bonnet term to the action, which does not affect the field equations in four dimensions. In order for the theory to admit a spin-2 degree of freedom with positive mass (around the Minkowski background), it is often required that ${m^2\equiv-1/\varkappa\beta>0}$. In the \textit{GR limit}, ${m\to\infty}$, the action reduces to the Einstein--Hilbert term, so we can also expect to get the GR solutions when this limit is applied to SG solutions. Since ${\varkappa f_{2,n-1}=-m^{-2}\delta_{n-1}^0}$, we can write the field equations for axially symmetric gyratons as (cf. \eqref{eq:fieldeqaxisym})
\begin{equation}\label{eq:feqSG}
\begin{aligned}
    \big(1-m^{-2}\triangle_1\big)\Phi_{21} &=\varkappa \Xi_{21}\;,
    \\
    \big(1-m^{-2}\triangle_0\big)\Phi_{22}-m^{-2}\diamondsuit\Phi_{21}&=\varkappa \Xi_{22}\;,
\end{aligned}
\end{equation}
where $\Phi_{21}$ and $\Phi_{22}$ are again given by \eqref{eq:Phiaxisym}.

\subsection{Homogeneous parts}

As before, let us start by identifying the homogeneous parts $J_{\textrm{hom}}$ and $H_{\textrm{hom}}$. This time, they obey the differential equations of the fourth order,
\begin{equation}
\begin{aligned}
    \big(1-m^{-2}\triangle_1\big){\circledcirc} J_{\textrm{hom}} &=0\;,
    \\
    \big(1-m^{-2}\triangle_0\big)\triangle_0 H_{\textrm{hom}} &=0\;.
\end{aligned}
\end{equation}
The general solutions of these two independent equations are given by linear combinations
\begin{equation}
\begin{aligned}
    J_{\textrm{hom}} &=c_1(u)  m\rho  I_1(m \rho )+c_2(u) m\rho  K_1(m \rho )+c_3(u)\rho ^2+c_4(u)\;,
    \\
    H_{\textrm{hom}} &=c_5(u) I_0(m \rho )+c_6(u) K_0(m \rho )+c_7(u)\log\rho +c_8(u)\;,
\end{aligned}
\end{equation}
with eight arbitrary functions $c_i(u)$. With this in hand we can now focus on vacuum solutions.

\subsection{Vacuum}

In contrast to GR gyratons, the gyratons in SG solving \eqref{eq:feqSG} may be of type III even in the region with no matter (${\bs{T}=0}$). The condition ${\Xi_{21}=0}$ is satisfied by ${J=J_{\textrm{hom}}}$ with arbitrary $c_i(u)$. In what follows, we focus on solutions that approach the vacuum GR solutions for ${\rho\to\infty}$ and give rise to continuous metric in Cartesian coordinates. These two assumptions lead to ${c_1(u)=c_3(u)=0}$ and ${c_2(u)=-c_4(u)}$, respectively. After renaming the function $c_4(u)$ to match \eqref{eq:vacuumsolgr}, we get
\begin{equation}\label{eq:JSG}
    J=\frac{\varkappa  \chi_{\textrm{J}}(u)}{4 \pi }\big(1-m\rho  K_1(m \rho )\big)\;.
\end{equation}
The corresponding curvature component
${\Phi_{21}}$ is non-zero,
\begin{equation}
    \Phi_{21}=\frac{i \varkappa  \chi_{\textrm{J}}(u) m^3}{2^4\sqrt{2}\pi}K_1(m \rho )\;,
\end{equation}
which signifies the algebraic type III as we foreshadowed.

Moving on to the second equation of \eqref{eq:feqSG} with ${\Xi_{22}=0}$, we observe that the particular part $H_{\textrm{part}}$ cannot be found by blindly following the methods presented in Section~\ref{ssec:axialsym} in this specific case. Namely, we cannot use the formula \eqref{eq:Phi22h},
\begin{equation}
    \Phi_{22}^{\textrm{part}} = \mathcal{H}_0^{-1}\left[\frac{\mathcal{H}_0\left[m^{-2}\diamondsuit\Phi_{21}\right](s)}{1+m^{-2}s^2}\right](\rho)\;,
\end{equation}
because the integral in the Hankel transform of the expression
\begin{equation}
    m^{-2}\diamondsuit\Phi_{21}=\frac{\varkappa^2 m^4  \chi_{\textrm{J}}^2(u)}{2^6 \pi^2 } \left(2 K_0(m \rho ){}^2+K_1(m \rho ){}^2\right)\equiv m^4 U(u,m\rho)
\end{equation}
does not converge. Despite this inconvenience, we can find $\Phi_{22}$ by direct integration, which results in
\begin{equation}
\begin{aligned}
    \Phi_{22} &=\frac{m^2}{2}c_5(u)I_0(m\rho)+ \frac{ m^2}{2}c_6(u)K_0(m\rho)+m^6\int_{\rho_1}^{\rho }\!\!d\tilde{\rho}\, \tilde{\rho}  \big(I_0(m \tilde{\rho})K_0(m \rho ) -I_0(m \rho ) K_0(m \tilde{\rho})\big)U(u,m\tilde{\rho})
    \\
    &=\frac{ m^2}{2}c_6(u)K_0(m\rho)+m^6\int_0^{\infty}\!\! d\tilde{\rho}\, \tilde{\rho}   \left[Z(m \tilde{\rho},m \rho)-\theta(\rho_1-\tilde{\rho})I_0(m \tilde{\rho})K_0(m \rho )\right]U(u,m\tilde{\rho})\;.
\end{aligned}
\end{equation}
In this derivation, we included two arbitrary functions $c_5(u)$ and $c_6(u)$ corresponding to the freedom in adding a homogeneous part. The function $c_5(u)$ was set to 
\begin{equation}
    c_5(u)=2m^4\int_{\rho_0}^{\infty}\!\!d\tilde{\rho}\, \tilde{\rho} K_0(m \tilde{\rho}) U(u,m\tilde{\rho})
\end{equation} 
so as to achieve vanishing $\Phi_{22}$ for ${\rho\to\infty}$. We also rewrote the expression in terms of the Heaviside step function $\theta$ and the function $Z$,
\begin{equation}
    Z(x,y)\equiv\begin{cases}
    I_0(x)K_0(y)\;, &x<y\;,
    \\
    I_0(y)K_0(x)\;, &x>y\;.
    \end{cases}
\end{equation}
The choice of $c_5(u)$ also guarantees the convergence of the convolution integral with the Green's function \eqref{eq:partHgf}, which we can use to find $H$, where without loss of generality we choose ${\rho_1=\rho_0}$. The corresponding solution is then given by
\begin{equation}\label{eq:HSG}
\begin{aligned}
    H &=c_6(u)\big(K_0(m\rho)+\log(\rho/\rho_0)\big)+2m^6\int_{0}^\infty\!\!d\rho'\,\rho'L(\rho,\rho')\int_0^{\infty}\!\! d\tilde{\rho}\, \tilde{\rho}   \left[Z(m \tilde{\rho},m \rho')-\theta(\rho_0-\tilde{\rho})I_0(m \tilde{\rho})K_0(m \rho' )\right]U(u,m\tilde{\rho})
    \\
    &\feq-\frac{\varkappa^2 m^2 \chi_{\textrm{J}}^2(u)}{2^6 \pi^2 }   \big(m^2 \rho^2  K_0(m \rho ){}^2+m \rho K_1(m \rho ) K_0(m \rho )-m^2 \rho^2  K_1(m \rho ){}^2+\log(\rho/\rho_0)\big)\;.
\end{aligned}
\end{equation}
The function $c_6(u)$ can be determined by comparing the asymptotic behavior of $H$ for ${\rho\to\infty}$ with \eqref{eq:vacuumsolgr}. To get a match, we set
\begin{equation}
    c_6(u)=-2m^6\int_{0}^\infty\!\!d\rho'\,\rho'\int_0^{\infty}\!\! d\tilde{\rho}\, \tilde{\rho}   \left[Z(m \tilde{\rho},m \rho')-\theta(\rho_0-\tilde{\rho})I_0(m \tilde{\rho})K_0(m \rho' )\right]U(u,m\tilde{\rho})+\frac{\varkappa^2 m^2  \chi_{\textrm{J}}^2(u)}{2^6 \pi^2 }+\frac{\varkappa\chi_{\textrm{H}}(u)}{2\pi}\;,
\end{equation}
which then gives us the same asymptotic expansion
\begin{equation}
    H(\rho\to\infty)\approx\frac{\varkappa\chi_{\textrm{H}}(u)}{4\pi}\log\left(\frac{\rho^2}{\rho_0^2}\right)\;.
\end{equation}

\begin{figure}
    \centering
    \includegraphics[width=0.48\textwidth]{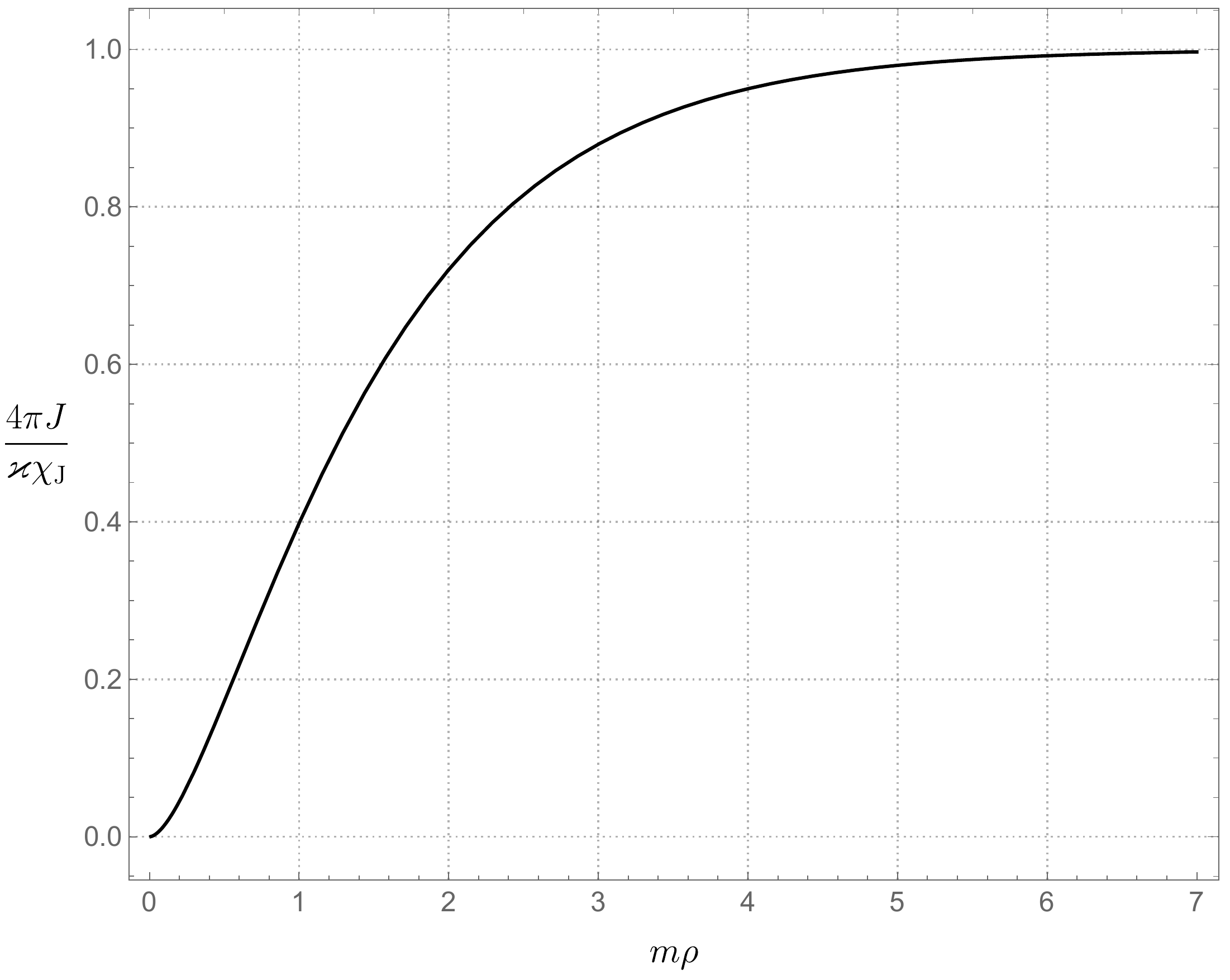}
    \quad
    \includegraphics[width=0.48\textwidth]{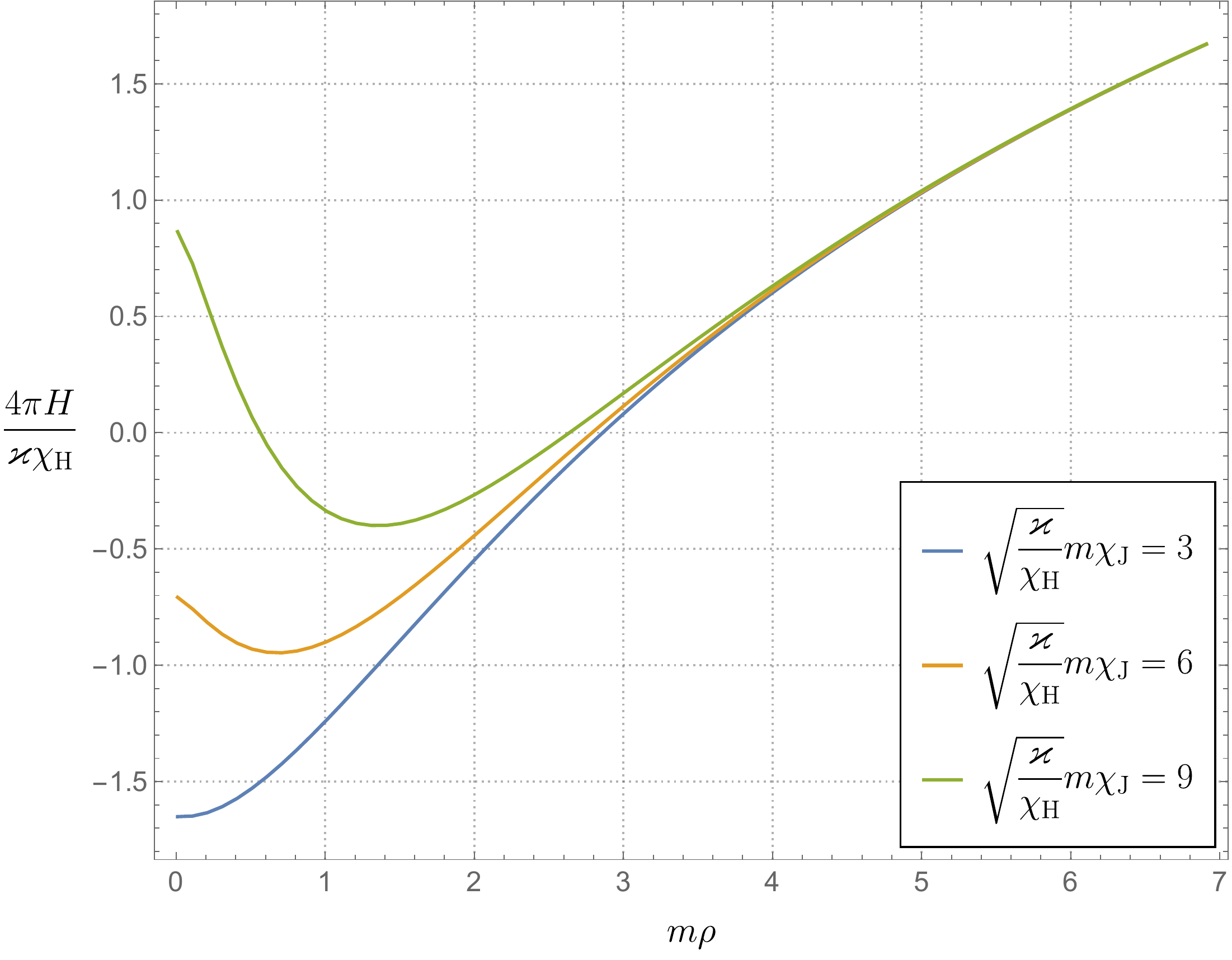}
    \caption{Vacuum gyraton in SG. Functions ${4\pi J}/{\varkappa\chi_{\textrm{J}}}$ (left) and function ${4\pi H}/{\varkappa\chi_{\textrm{H}}}$ (right) with respect to the variable $m\rho$ for the values: ${m\rho_0=3}$ and ${(\varkappa/\chi_{\textrm{H}})^{1/2}m\chi_{\textrm{J}}=3,6,9}$.}
    \label{fig:vac}
\end{figure}

Employing the properties of the Bessel functions, it is not difficult to check that the GR limit of the solution given by \eqref{eq:JSG} and \eqref{eq:HSG} is nothing but GR gyraton in vacuum \eqref{eq:vacuumsolgr}. Interestingly, we can also observe the regular behavior ${J=O(\rho^2)}$ and ${H=O(1)}$ near ${\rho=0}$, even though no regularization of the source was imposed by hand. In the slowly-rotating linearized regime $H$ reduces to
\begin{equation}
    H_{\textrm{lin}} =\frac{\varkappa\chi_{\textrm{H}}(u)}{4\pi}\left[\log\left(\frac{\rho^2}{\rho_0^2}\right)+2K_0(m\rho)\right]\;.
\end{equation}
The relevant graphs for the SG vacuum gyraton are plotted in dimensionless quantities in Figure~\ref{fig:vac}.

\section{Gyratons in IDG}\label{sec:gyridg}

Consider a non-local gravity with the form-factors \cite{Biswas:2011ar}:
\begin{equation}
    \varkappa\mathcal{F}_2(\square)=-4\varkappa\mathcal{F}_1(\square)=\frac{\mathcal{A}(\square)-1}{\square}\;, \quad \mathcal{F}_3(\square)=0\;,
\end{equation}
where $\mathcal{A}$ is an arbitrary analytic \textit{non-polynomial} function with no zeros in the complex plane satisfying ${\mathcal{A}(0)=1}$. This theory is often referred to as the infinite derivative gravity. The choice of the form-factor ensures that (around the Minkowski background) the theory has no ghosts or extra degrees of freedom when compared to GR. As a simple example, we take the exponential operator
\begin{equation}\label{eq:exp}
    \mathcal{A}(\square)=e^{-\ell^2\square}\;,
\end{equation}
which implies ${\varkappa f_{2,n-1}=(-\ell^2)^{n}/n!}$. The parameter $\ell$ is called the \textit{(length) scale of non-locality}. Einstein--Hilbert action is recovered in the \textit{(local) GR limit}, ${\ell\to 0}$. The exponential operator \eqref{eq:exp} has also a technical advantage over other common choices. It allows us to simplify the infinite double-sum operator in \eqref{eq:fieldeqaxisym},
\begin{equation}
    \sum_{n=1}^{\infty}\tfrac{(-\ell^2)^{n}}{n!}\sum_{k=0}^{n-1}\triangle_0^k\diamondsuit\triangle_1^{n-k-1}=\sum_{k=0}^{\infty}\triangle_0^k\diamondsuit\sum_{l=0}^{\infty} \tfrac{(-\ell^2)^{k+l+1}}{(k+l+1)!}\triangle_1^l=-\ell^2\int_0^1\!\! dt\, e^{-t\ell^2\triangle_0}\diamondsuit e^{-(1-t)\ell^2\triangle_1}\;,
\end{equation}
using the integral identity
\begin{equation}
    \int_0^1\!\! dt\,t^k (1-t)^l=\frac{k!l!}{(k+l+1)!}\;.
\end{equation}
This mathematical trick brings the field equations to much more tractable form with an integral instead of infinite double-sums,
\begin{equation}
\begin{aligned}
    e^{-\ell^2\triangle_1}\Phi_{21} &=\varkappa \Xi_{21}\;,
    \\
    e^{-\ell^2\triangle_0}\Phi_{22}-\ell^2\int_0^1\!\! dt\, e^{-t\ell^2\triangle_0}\diamondsuit e^{-(1-t)\ell^2\triangle_1}\Phi_{21}&=\varkappa \Xi_{22}\;.
\end{aligned}
\end{equation}
Let us now proceed to discuss the solutions of these non-local equations.

\subsection{Homogeneous parts}
Once more we start with homogeneous parts $J_{\text{hom}}$ and $H_{\text{hom}}$, which now obey the differential equations
\begin{equation}\label{eq:nlhom}
\begin{aligned}
    e^{-\ell^2\triangle_1}{\circledcirc} J_{\textrm{hom}} &=0\;,
    \\
    e^{-\ell^2\triangle_0}\triangle_0 H_{\textrm{hom}} &=0\;.
\end{aligned}
\end{equation}
It is a well-known fact \cite{Barnaby:2007ve,Barnaby:2008tc} that the structure of solutions of homogeneous equations is affected only by the operators with zeros in the complex plane. Following this rule, the non-local exponential operators should not change the homogeneous parts, thus the solutions of \eqref{eq:nlhom} should be equivalent to the solutions of \eqref{eq:JHhomeq}, which are given by \eqref{eq:JHhomgr}. However, this is only true for ${\rho>0}$ and we have no reasons to exclude the origin ${\rho=0}$ from the domain of functions on which the operators act. At this moment, we recall the result \eqref{eq:homdistrgr}, which states that ${J_{\textrm{hom}}=c_2(u)}$ as well as ${H_{\textrm{hom}}=c_3(u)\log\rho}$ are not homogeneous solutions when ${\rho=0}$ is taken into account in a distributional sense. Since the actions of $e^{-\ell^2\triangle_1}$ and $e^{-\ell^2\triangle_0}$ on Dirac-delta sources \eqref{eq:homdistrgr} are not even mathematically well-defined (the integrals in the Fourier space blow up), we are forced to set ${c_2(u)=c_3(u)=0}$, i.e.,
\begin{equation}\label{eq:IDGhom}
\begin{aligned}
    J_{\textrm{hom}} &=c_1(u)\rho^2\;,
    \\
    H_{\textrm{hom}} &=c_4(u)\;.
\end{aligned}
\end{equation}
For this reason, we should again expect the full solution to behave like ${J=O(\rho^2)}$ and ${H=O(1)}$ near ${\rho=0}$.

\subsection{Gaussian beam}
Let us consider the source \eqref{eq:gausssource} describing the spinning null matter of Gaussian-type distribution. This kind of source is not only physically relevant, but also makes the calculations exceptionally simple. The reason is because the formulas for the particular parts \eqref{eq:Phi21h} and \eqref{eq:Phi22h} reduce to
\begin{equation}\label{eq:Phi21hidg}
    \Phi_{21}^{\textrm{part}} = \varkappa\mathcal{H}_1^{-1}\left[e^{-\ell^2 \tilde{s}^2} \mathcal{H}_1[\Xi_{21}](\tilde{s})\right](\rho)
\end{equation}
and
\begin{equation}\label{eq:Phi22hidg}
    \Phi_{22}^{\textrm{part}} = \varkappa\mathcal{H}_0^{-1}\left[e^{-\ell^2 s^2} \mathcal{H}_0[\Xi_{22}](s)\right](\rho)+\ell^2 \int_0^1\!\! dt\,\mathcal{H}_0^{-1}\left[e^{-(1-t)\ell^2 s^2}\mathcal{H}_0\left[\diamondsuit\mathcal{H}_1^{-1}\left[e^{(1-t)\ell^2\tilde{s}^2}\mathcal{H}_1[\Phi_{21}](\tilde{s})\right](\rho)\right](s)\right](\rho)\;.
\end{equation}
respectively. Although these expressions may look pretty intimidating at first sight, the actual calculations will involve just repeated evaluation of Hankel transforms of Gaussian-type functions (Gaussian functions multiplied by even/odd polynomials) which are then turned into other Gaussian-type functions. Moreover, because all integrals converge, we can follow the methods from Section~\ref{ssec:axialsym} exactly as written there.

In particular, the formula \eqref{eq:Phi21hidg} leads to
\begin{equation}
    \Phi_{21}^{\textrm{part}} =\frac{i \varkappa  \chi_{\textrm{J}}(u) }{2^6 \sqrt{2} \pi  \left(\ell ^2+\epsilon^2 \right)^2}\rho  e^{-\frac{\rho ^2}{4 \left(\ell ^2+\epsilon ^2\right)}}\;.
\end{equation}
which can be then integrated using \eqref{eq:partJcf} to obtain
\begin{equation}\label{eq:JIDG}
    J=\frac{\varkappa  \chi_{\textrm{J}}(u)}{4 \pi }\Big(1-e^{-\frac{\rho ^2}{4 \left(\ell ^2+\epsilon ^2\right)}}\Big)\;,
\end{equation}
where we used the freedom in adding a homogeneous part proportional to $\rho^2$ that provides GR-like asymptotic behavior for ${\rho\to\infty}$, cf. \eqref{eq:JgausGR}. Notice that the difference lies only in the replacement of $\epsilon^2$ by the effective width~${\epsilon^2{+}\ell^2}$.

With this $J$ in hand, we can now express the operator $\diamondsuit$, and evaluate all the Hankel transforms in~\eqref{eq:Phi22hidg},
\begin{equation}\label{eq:phi22gaussnl}
    \Phi_{22}^{\textrm{part}} =\frac{\varkappa\chi_{\textrm{H}}(u) }{2^3 \pi  \left(\ell ^2+\epsilon ^2\right)}e^{-\frac{\rho ^2}{4 \left(\ell ^2+\epsilon ^2\right)}}+\frac{\varkappa^2  \chi_{\textrm{J}}^2(u) }{2^8 \pi ^2 \epsilon ^2\left(2\ell ^2+\epsilon ^2\right)}e^{-\frac{\rho ^2}{2 (2\ell ^2+\epsilon ^2)}}+ \frac{\varkappa ^2 \chi_{\textrm{J}}^2(u) \ell^2}{2^{10} \pi ^2 }\int_0^1\!\!dt\,\left[\alpha_0(t)+\alpha_2(t)\rho^2\right]e^{-\beta(t)\rho ^2}\;,
\end{equation}
where we introduced the functions
\begin{equation}
\begin{aligned}
    \alpha_0(t) &\equiv -\frac{4 (3+t (2-(4-t) t)) \ell ^6+4 (11-t (t+4)) \ell ^4 \epsilon ^2+12 (3-t) \ell ^2 \epsilon ^4+8 \epsilon ^6}{\left[(1+(1-t)t) \ell ^4+(3-t) \ell ^2 \epsilon ^2+\epsilon ^4\right]^3}\;,
    \\
    \alpha_2(t) &\equiv \frac{\left(\ell ^2+\epsilon ^2\right) \left((t+2) \ell ^2+3 \epsilon ^2\right)}{\left[(1+(1-t)t) \ell ^4+(3-t) \ell ^2 \epsilon ^2+\epsilon ^4\right]^3}\;,
    \\
    \beta(t) &\equiv\frac{ (t+1) \ell ^2+2 \epsilon ^2}{4 \left[(1+(1-t) t) \ell ^4+(3-t) \ell ^2 \epsilon ^2+ \epsilon ^4\right]}\;.
\end{aligned}
\end{equation}
We can observe that the first two terms and the integrand in the third term of \eqref{eq:phi22gaussnl} as well as the term
\begin{equation}
    \frac{\big(J_{,\rho}\big)^2}{2 \rho^2}= \frac{\varkappa ^2 \chi_{\textrm{J}}^2(u) }{128 \pi ^2 \left(\ell ^2+\epsilon ^2\right)^2}e^{-\frac{\rho ^2}{2 \left(\ell ^2+\epsilon ^2\right)}}
\end{equation}
have the same form,
\begin{equation}
    W_{\varsigma}=\big(a_0(u)+a_2(u)\rho^2\big)e^{-b\rho^2}=\big(a_0(u)-a_2(u)\pp_{b}\big)e^{-b\rho^2}\;.
\end{equation}
Here $\varsigma$ labels the individual `summands', where the integrand is also treated as a summand with a continuous index~$t$. Therefore, we can formally write the right-hand side of \eqref{eq:Poisson} as
\begin{equation}
    W=\sumint\nolimits_{\varsigma}\!\!\! W_{\varsigma}\;.
\end{equation}
Owing to the linearity of \eqref{eq:Poisson}, we can calculate the solutions for individual terms by means of the Green's function~\eqref{eq:partHgf},
\begin{equation}
\begin{aligned}
    H_{\varsigma} &=G\star W_{\varsigma}=G\star \big(a_0(u)-a_2(u)\pp_{b}\big)e^{-b\rho^2}=\big(a_0(u)-a_2(u)\pp_{b}\big)\big(G\star e^{-b\rho^2}\big)
    \\
    &=\frac{a_0(u) b+a_2(u)}{4 b^2}\left[\log \left(\rho^2/\rho_0^2\right)-\Ei\left(-b \rho ^2\right)\right]+\frac{a_2(u)}{4b^2}e^{-b \rho ^2}\;,
\end{aligned}
\end{equation}
where we used again \eqref{eq:convgauss}. After summing and integrating these terms we arrive at the result
\begin{equation}\label{eq:HIDG}
\begin{aligned}
    H &=\frac{\varkappa\chi_{\textrm{H}}(u)}{4 \pi }  \left[\log \left(\frac{\rho ^2}{\rho_0^2}\right)-   \Ei\left(-\frac{\rho ^2}{4 \left(\ell ^2+\epsilon ^2\right)}\right)\right]
    \\
    &\feq+\frac{\varkappa^2  \chi_{\textrm{J}}^2(u) }{2^8 \pi ^2}  \left\{\frac{\ell ^2}{ \epsilon ^2 \left(\ell ^2+2\epsilon ^2\right)} \log \left(\frac{\rho ^2}{\rho_0^2}\right)+\frac{1}{\ell^2+\epsilon^2}\Ei\left(-\frac{\rho ^2}{2 \left(\ell ^2+\epsilon ^2\right)}\right)-\frac{1}{\epsilon^2}\Ei\left(-\frac{\rho ^2}{2(2 \ell ^2+\epsilon ^2)}\right)\right.
    \\
    &\feq+\int\limits_0^1 dt\;\frac{ \ell ^2}{ \left(  (t+1) \ell ^2+2   \epsilon ^2\right)^2} \left[\Ei\left(-\frac{(1+t) \ell ^2+2 \epsilon ^2}{4 \left[((1-t) t+1) \ell ^4+(3-t) \ell ^2 \epsilon ^2+\epsilon ^4\right]}\rho^2\right)\right.
    \\
    &\feq\left.\left.+\frac{\left(\ell ^2+\epsilon ^2\right) \left((t+2) \ell ^2+3 \epsilon ^2\right) }{((1-t) t+1) \ell ^4+(3-t) \ell ^2 \epsilon ^2+\epsilon ^4}\exp\left(-\frac{(1+t) \ell ^2+2 \epsilon ^2}{4 \left[((1-t) t+1) \ell ^4+(3-t) \ell ^2 \epsilon ^2+\epsilon ^4\right]}\rho^2\right)\right]\right\}\;.
\end{aligned}
\end{equation}

\begin{figure}
    \centering
    \includegraphics[width=0.48\textwidth]{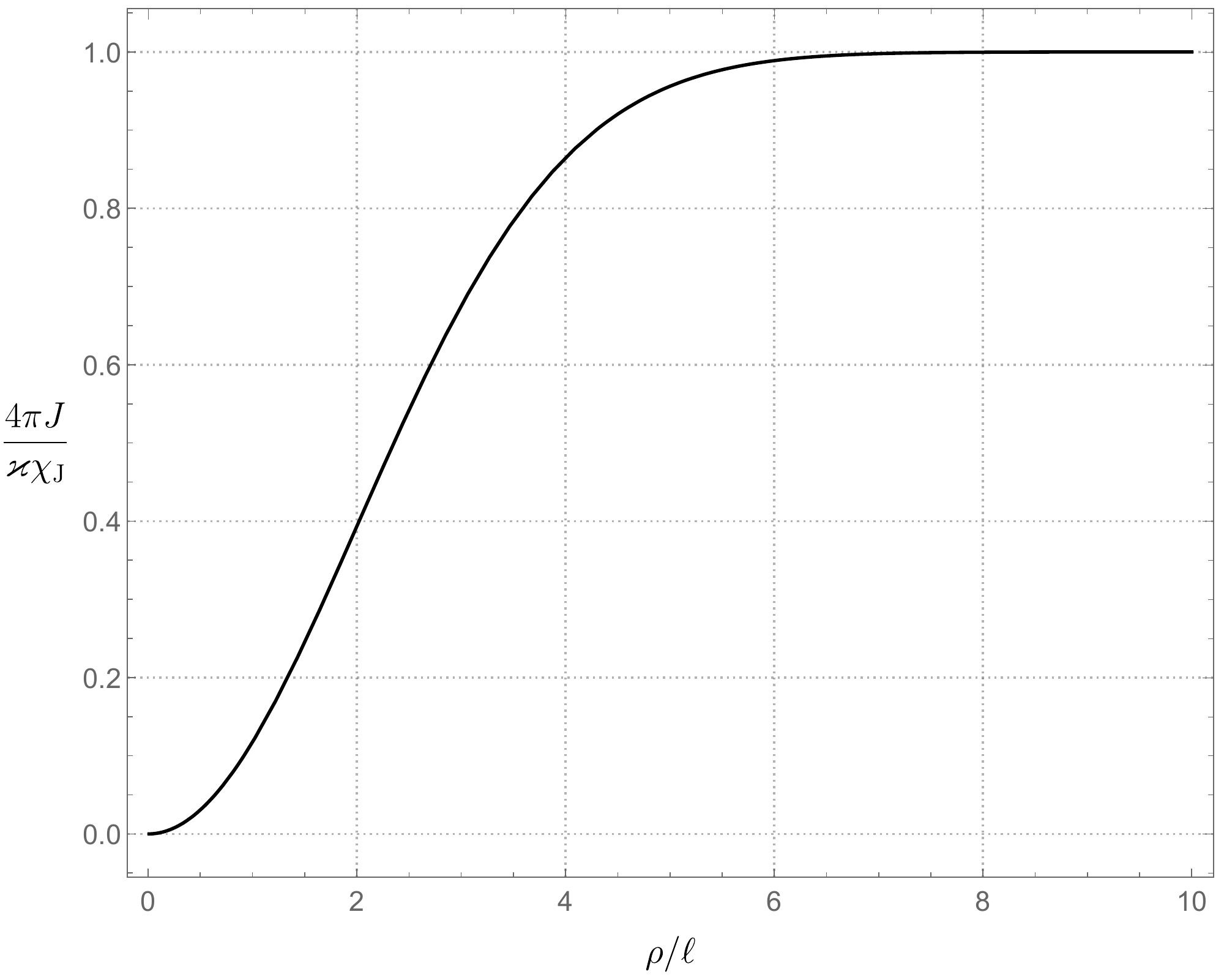}
    \quad
    \includegraphics[width=0.48\textwidth]{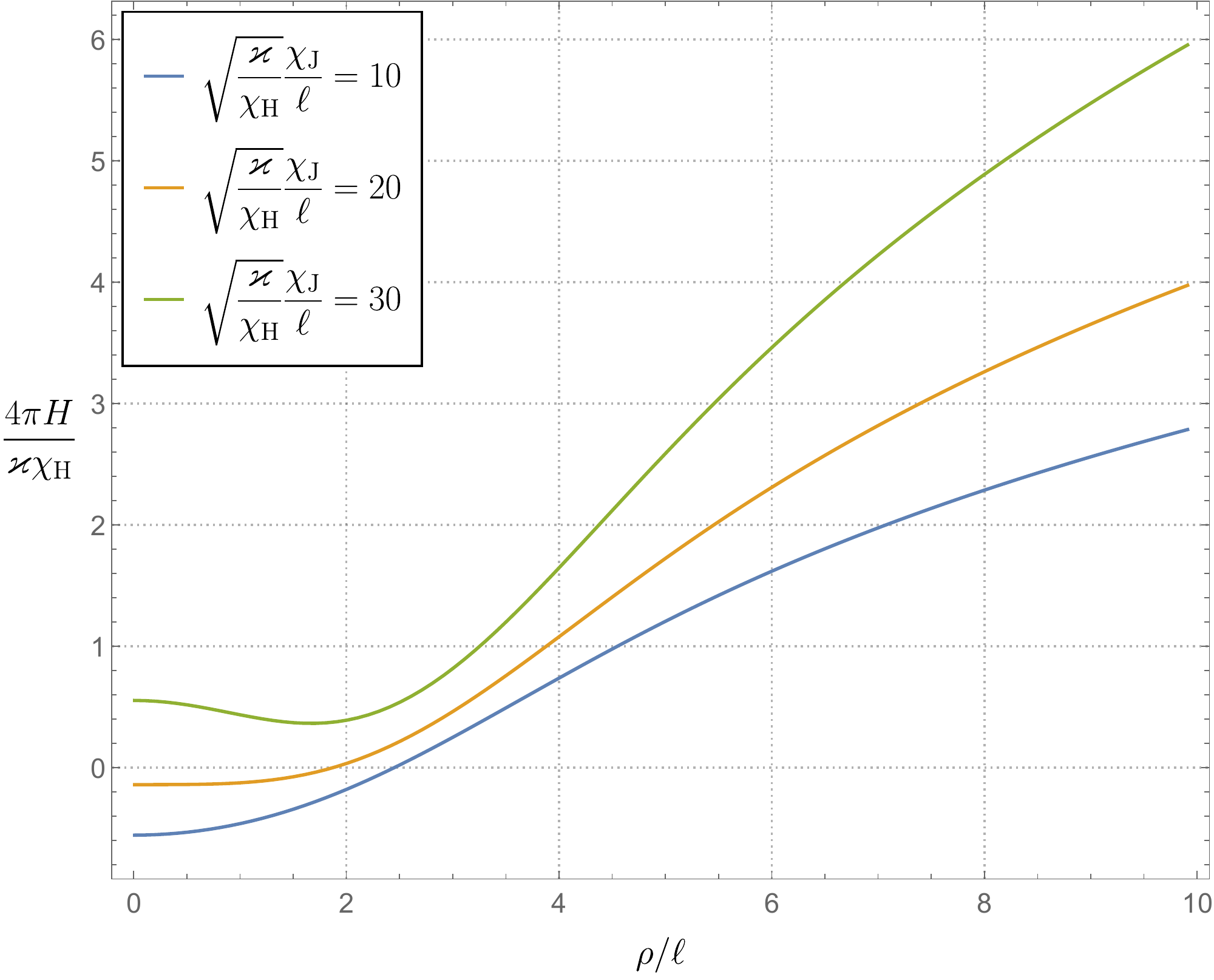}
    \caption{Gaussian beam gyraton in IDG. Functions ${4\pi J}/{\varkappa\chi_{\textrm{J}}}$ (left) and function ${4\pi H}/{\varkappa\chi_{\textrm{H}}}$ (right) with respect to the variable $\rho/\ell$ for the values: ${\epsilon/\ell=1}$, ${\rho_0/\ell=3}$, and ${(\varkappa/\chi_{\textrm{H}})^{1/2}\chi_{\textrm{J}}/\ell=10,20,30}$.}
    \label{fig:gauss}
\end{figure}

Notice that the asymptotic behavior of $H$ for ${\rho\to\infty}$,
\begin{equation}\label{eq:HasympIDG}
    H(\rho\to\infty) \approx\left[\frac{   \varkappa \chi_{\textrm{H}}(u)}{4 \pi }+\frac{\varkappa^2  \chi_{\textrm{J}}^2(u)}{2^8 \pi ^2}\frac{ \ell^2}{\epsilon ^2 \left(\ell ^2+2\epsilon ^2\right)}\right]\log \left(\frac{\rho ^2}{\rho_0^2}\right)\;,
\end{equation}
is governed by a different constant compared to the GR solution. In IDG, unlike in GR or SG, this logarithmic behavior cannot be modified by adding a homogeneous part, cf. \eqref{eq:IDGhom}. If we take the GR limit of the non-local solution given by \eqref{eq:JIDG} and \eqref{eq:HIDG} and use the properties of the exponential integral, we recover the Gaussian beam solution in GR, \eqref{eq:JgausGR} and \eqref{eq:HgausGR}. Near ${\rho=0}$ the solution has the expected behavior ${J=O(\rho^2)}$ and ${H=O(1)}$. In the linearized approximation of slow rotation, $H$ simplifies to
\begin{equation}
    H_{\textrm{lin}} =\frac{\varkappa\chi_{\textrm{H}}(u)}{4 \pi }  \left[\log \left(\frac{\rho ^2}{\rho_0^2}\right)-   \Ei\left(-\frac{\rho ^2}{4 \left(\ell ^2+\epsilon ^2\right)}\right)\right]\;,
\end{equation}
where we also see the effective replacement of $\epsilon^2$ by ${\epsilon^2+\ell^2}$ when compared to the GR result. If we also take the limit ${\epsilon\to 0}$, the geometry reduces to the gyratons solution in the linearized IDG \cite{Boos:2020ccj}. Graphs of the IDG solution for the Gaussian beam in dimensionless quantities is depicted in Figure~\ref{fig:gauss}.

\section{Higher-order curvature gravity}\label{sec:highordcurv}

In this section, we would like to briefly comment on gyraton solutions of theories that are of the cubic and higher orders in curvature. 
First, let us point out that Lagrangians of the $n$th order in curvature (with no terms of a lower order than $n$) lead to field equations which are at least of the ${(n{-}1)}$th order in curvature. This can be seen by noticing that any Lagrangian $L = L(\bs{g}, \bs{\nabla}{\cdots}\bs{\nabla}\bs{R})$ can be rearranged to the form that is totally symmetric in derivatives,
\begin{equation}\label{eq:Ltotsym}
    L = L(\bs{g}, \bs{\nabla}_{\bs{a}_1} \bs{R}, \bs{\nabla}_{(\bs{a}_1}\bs{\nabla}_{\bs{a}_2)}\bs{R},\ldots, \bs{\nabla}_{(\bs{a}_1}{\cdots}\bs{\nabla}_{\bs{a}_p)}\bs{R})\;,
\end{equation}
because we can get rid of the anti-symmetric parts using the commutator at the expense of introducing one more curvature tensor. The field equations derived from the Lagrangian \eqref{eq:Ltotsym} then read \cite{Iyer:1994ys}
\begin{equation}
    \frac{\partial L}{\partial \bs{g_{ab}}} + \bs{E^a{}_{cde}R^{bcde}} + 2 \bs{\nabla_c \nabla_d E^{cabd}} + \frac12 \bs{g^{ab}} L=\bs{T^{ab}} \;,
\end{equation}
where we denoted
\begin{equation}
    \bs{E^{abcd}} \equiv \frac{\partial L}{\partial\bs{R_{abcd}}} - \bs{\nabla}_{\bs{a}_1} \frac{\partial L}{\partial\bs{\nabla}_{\bs{a}_1} \bs{R_{abcd}}} + \ldots + (-1)^p \bs{\nabla}_{\bs{(}\bs{a}_1}{\cdots} \bs{\nabla}_{\bs{a}_p\bs{)}} \frac{\partial L}{\partial\bs{\nabla}_{\bs{(}\bs{a}_1}{\cdots}\bs{\nabla}_{\bs{a}_p\bs{)}} \bs{R_{abcd}}}\;.
\end{equation}
Therefore, starting with a Lagrangian involving scalar curvature invariants of $n$th order and higher, only $\bs{\nabla_c\nabla_d E^{cabd}}$ contains terms of a lower order in curvature, namely ${n{-}1}$.

As mentioned in Section \ref{ssec:typeIIIppwaves}, rank-2 tensors that are cubic or of a higher order in curvature vanish and hence we can conclude that the only contributions to the field equations for gyratons following from Lagrangians that are cubic in curvature must be quadratic in curvature; moreover, they are given by $\bs{\nabla_c\nabla_d E^{cabd}}$. Scalar curvature invariants of the fourth and higher orders do not contribute at all.

However, a generic Lagrangian cubic in curvature cannot be recast in a form involving only the wave operator $\square$ using the integration by parts, Bianchi identities, and the symmetry of the Riemann tensor as one can do in the quadratic case \eqref{eq:action}.\footnote{The contraction of two covariant derivatives gives $\square$, no matter where they appear in the cubic term $\bs{\nabla}{\cdots}\bs{\nabla R \nabla}{\cdots}\bs{\nabla R \nabla}\cdots\bs{\nabla R}$ of the action that is totally symmetric in derivatives because either $\bs{\nabla}{\cdots}\bs{\nabla \nabla_a\nabla}{\cdots}\bs{\nabla \nabla^a \nabla}{\cdots}\bs{\nabla R} = \square\bs{\nabla}{\cdots}\bs{\nabla R}$ or we have the term of the form $\bs{\nabla}{\cdots}\bs{\nabla \nabla_a \nabla}{\cdots}\bs{\nabla R \nabla}{\cdots}\bs{\nabla \nabla^a \nabla}{\cdots}\bs{\nabla R \nabla}\cdots\bs{\nabla R}$, where we can employ $\bs{\nabla_c t}_1 \bs{\nabla^c t}_2 \bs{t}_3 = \frac12 \bs{t}_1 \bs{t}_2 \square\bs{t}_3 - \frac12 \square\bs{t}_1 \bs{t}_2 \bs{t}_3 - \frac12 \bs{t}_1 \square\bs{t}_2 \bs{t}_3$. The last identity follows from the combination of $\square(\bs{t}_1 \bs{t}_2) \bs{t}_3 = - \nabla_c(\bs{t}_1 \bs{t}_2) \nabla^c\bs{t}_3 = (\bs{t}_1 \bs{t}_2) \square\bs{t}_3$ (integrations by parts) and $\square(\bs{t}_1 \bs{t}_2) \bs{t}_3 = \square\bs{t}_1 \bs{t}_2 \bs{t}_3 + 2 \nabla_c\bs{t}_1 \nabla^c\bs{t}_2 \bs{t}_3 + \bs{t}_1 \square\bs{t}_2 \bs{t}_3$ (Leibniz rule). A covariant derivative contracted with the Riemann tensor on which it is applied can be eliminated using the contracted Bianchi identities $\bs{\nabla^d R_{abcd}} = 2 \bs{\nabla_{[a}R_{b]c}}$, $\bs{\nabla^b R_{ab}} = \tfrac12 \bs{\nabla_a}R$. Unfortunately, it seems that there is no appropriate way of simplifying the contractions containing $\bs{\nabla}{\cdots}\bs{\nabla \nabla^a \nabla}{\cdots}\bs{\nabla R} \bs{\nabla}{\cdots}\bs{\nabla R_{abcd}}$ in the cubic terms.} We are thus not able to explicitly express $\bs{\nabla_c\nabla_d E^{cabd}}$ in a compact form. Nevertheless, the results of Section \ref{ssec:TensorsQuadraticInCurvature} imply that the only possible non-vanishing contributions to the field equations following from Lagrangians cubic in curvature are of the forms
\begin{equation}\label{eq:fromcubic}
\begin{aligned}
    &\bs{\nabla}{\cdots}\bs{\nabla S}_{\circ\bullet} \bs{\nabla}{\cdots}\bs{\nabla S}_{\circ\bullet}\;,
    \\
    &\bs{\nabla}{\cdots}\bs{\nabla S}_{\circ\bullet} \bs{\nabla}{\cdots}\bs{\nabla C}_{\circ\bullet\bullet\bullet}\;,
    \\
    &\bs{\nabla}{\cdots}\bs{\nabla S}_{\bullet\bullet} \bs{\nabla}{\cdots}\bs{\nabla C}_{\circ\bullet\circ\bullet}\;,
    \\
    &\bs{\nabla}{\cdots}\bs{\nabla C}_{\circ\bullet\bullet\bullet} \bs{\nabla}{\cdots}\bs{\nabla C}_{\circ\bullet\bullet\bullet}\;,
    \\
    &\bs{\nabla}{\cdots}\bs{\nabla C}_{\circ\bullet\circ\bullet} \bs{\nabla}{\cdots}\bs{\nabla C}_{\bullet\bullet\bullet\bullet}\;,
\end{aligned}
\end{equation}
where $\circ$ and $\bullet$ represent free and dummy indices, respectively. All such rank-2 tensors have only b.w.\ $-2$ components constructed from two b.w.\ $-1$ components of the TF Ricci and Weyl tensors. In other words, the first equation of the field equations for gyraton metrics \eqref{eq:fieldeqgyr} (equation for $J$, independent of $H$) is unaffected by higher-order terms in the action, so the solution for $J$ remains the same. On the other hand, the second equation of \eqref{eq:fieldeqgyr} (equation for $H$) is modified by additional terms that are quadratic in $\Phi_{21}$ (i.e., quadratic in $J$).


\section{Conclusions}\label{sec:C}

In this work, we derived field equations for gyratons in generic theories of gravity that are quadratic in curvature and contain an arbitrary number of covariant derivatives. We also commented on theories with higher order terms in curvature. Since the gyraton metric \eqref{eq:gyrmetric} belongs to the pp-waves of type III, many terms in the field equations either vanish or get drastically simplified, as shown in Section~\ref{sec:typeIIIppwaves}. The resulting set of two partial differential equations \eqref{eq:fieldeqgyr} is partly linear and decoupled, which makes the system completely solvable using standard methods for linear differential equations. In particular, for the axially symmetric case, we show that these equations become ordinary differential equations \eqref{eq:fieldeqaxisym}, and can be solved by means of the Hankel transforms, Green's function, etc. The reduced field equations \eqref{eq:fieldeqgyr} and \eqref{eq:fieldeqaxisym}, and methods presented in Section~\ref{ssec:axialsym} are readily applicable to various theories of gravity (provided that Hankel transforms exists and convolution integrals converge).

To demonstrate the application of our equations, we considered Stelle's fourth derivative gravity and the non-local gravity with an infinite number of derivatives. In the former (SG), we found a vacuum gyraton solution that is described by functions \eqref{eq:JSG} and \eqref{eq:HSG}, see Figure \ref{fig:vac}. In the latter (IDG), we obtained the gyraton solution for the Gaussian beam, given by the functions \eqref{eq:JIDG} and \eqref{eq:HIDG}, see Figure \ref{fig:gauss}. The functions $J$ and $H$ are regular in both cases. Furthermore, the obtained solutions reduce to the corresponding gyratons in GR when we take the GR limits of the theories. SG vacuum gyraton also approaches GR vacuum gyraton far from the source, while IDG Gaussian beam gyraton shows different logarithmic behavior then GR Gaussian beam gyraton, see \eqref{eq:HasympIDG}. This is most likely a consequence of the fact the non-locality plays an important role only near the sources. The Gaussian-type source extends to infinity while the source that generates the vacuum SG solution is located at ${\rho=0}$.

Since all pp-waves of type III are of VSI, all gyratons within this class are always free of scalar curvature singularities. To decide on the presence/absence of the non-scalar curvature singularities \cite{Geroch1968,Ellis1977} one needs to investigate the components of the curvature in PP frames along all timelike and null geodesics. However, this is a rather non-trivial task that deserves a proper investigation in a separate project.

One natural continuation of our research was hinted in Section~\ref{sec:highordcurv}, i.e., the generalization to completely generic actions that are analytic in $\bs{R}$ and $\bs{\nabla}$. As mentioned there, the actions that are of quartic and higher orders in the curvature cannot contribute to the field equations of gyratons. Thus, since we already dealt with the contributions from the quadratic terms, the only remaining terms that must be worked out are of the cubic order in curvature. It was also pointed out above that their contributions to the field equations should be of the form \eqref{eq:fromcubic}, which can only affect the function $H$ while the function $J$ remains unchanged.

Another interesting project is also generalization to spacetimes with constant non-zero Ricci scalar $R$, which could be interpreted as gyratons (spinning null sources) propagating in (anti-)de Sitter background. Although, this line of research requires considering type II spacetimes with non-zero components of b.w.~0, the reduction of equation might still be significant because the b.w.~0 are necessarily constant.


\section*{Acknowledgements}

We would like to thank to Jens Boos (Williamsburg, US) for useful discussions. I.K. was supported by Netherlands Organization for Scientific Research (NWO) grant no. 680-91-119. T.M. acknowledges the support of the Czech Science Foundation GA\v{C}R grant no. GA19-09659S. The works of E.K. and S.D. are supported by the TUBITAK grant no. 119F241.


\appendix


\section{NP formalism}\label{apx:NP}

In this appendix, we gather the most important formulas of the Newman--Penrose formalism; for more details we refer the reader to \cite{Stephani:2003tm}. The formalism makes use of the orthonormal null covector frame ${\bs{l}}$, ${\bs{n}}$, ${\bs{m}}$, and ${\bar{\bs{m}}}$ satisfying \eqref{eq:orthonormalframe}. In this frame, the covariant derivative $\bs{\nabla}$ can be expressed by means of the \textit{directional derivatives} $\DD$, $\nDelta$, $\delta$, and~$\bar\delta$,
\begin{equation}\label{eq:nablaDDeltadeltadelta}
    \bs{\nabla} = - \bs{n} \DD - \bs{l} \nDelta + \bar{\bs{m}} \delta + \bs{m} \bar\delta\;.
\end{equation}
Derivatives of the frame vectors are characterized using 12 complex functions called the \textit{spin coefficients} commonly denoted by lower-case Greek letters:
\begin{equation}\label{eq:NPspincoeffs}
\begin{aligned}
    \DD\bs{l} &= (\varepsilon + \bar\varepsilon) \bs{l} - \bar\kappa \bs{m} - \kappa \bar{\bs{m}}\;,
    &
    \DD \bs{n} &= -(\varepsilon + \bar\varepsilon) \bs{n} + \pi \bs{m} + \bar\pi \bar{\bs{m}}\;,
    &
    \DD \bs{m} &= \bar\pi \bs{l} - \kappa \bs{n} + (\varepsilon - \bar\varepsilon) \bs{m}\;,
    \\
    \nDelta \bs{l} &= (\gamma + \bar\gamma) \bs{l} - \bar\tau \bs{m} - \tau \bar{\bs{m}}\;,
    &
    \nDelta \bs{n} &= -(\gamma + \bar\gamma) \bs{n} + \nu \bs{m} + \bar\nu \bar{\bs{m}}\;,
    &
    \nDelta \bs{m} &= \bar\nu \bs{l} - \tau \bs{n} + (\gamma - \bar\gamma) \bs{m}\;,
    \\
    \delta \bs{l} &= (\bar\alpha + \beta) \bs{l} - \bar\rho \bs{m} - \sigma \bar{\bs{m}}\;,
    &
    \delta \bs{n} &= -(\bar\alpha + \beta) \bs{n} + \mu \bs{m} + \bar\lambda \bar{\bs{m}}\;,
    \\
    \delta \bs{m} &= \bar\lambda \bs{l} - \sigma \bs{n} - (\bar\alpha - \beta) \bs{m}\;,
    &
    \bar\delta \bs{m} &= \bar\mu \bs{l} - \rho \bs{n} + (\alpha - \bar\beta) \bs{m}\;.
\end{aligned}
\end{equation}
When acting on scalars, the commutators of directional derivatives read:
\begin{equation}\label{eq:commutonscal}
\begin{aligned}
[\nDelta, \DD] &=(\gamma+\bar{\gamma}) \DD+(\varepsilon+\bar{\varepsilon}) \nDelta-(\tau+\bar{\pi}) \bar{\delta}-(\bar{\tau}+\pi) \delta\;,
\\
[\delta, \DD] &=(\bar{\alpha}+\beta-\bar{\pi}) \DD+\kappa \nDelta-\sigma \bar{\delta}-(\bar{\rho}+\varepsilon-\bar{\varepsilon}) \delta\;,
\\
[\delta, \nDelta] &=-\bar{\nu} \DD+(\tau-\bar{\alpha}-\beta) \nDelta+\bar{\lambda} \bar{\delta}+(\mu-\gamma+\bar{\gamma}) \delta\;,
\\
[\bar{\delta}, \delta] &=(\bar{\mu}-\mu) \DD+(\bar{\rho}-\rho) \nDelta-(\bar{\alpha}-\beta) \bar{\delta}-(\bar{\beta}-\alpha) \delta\;.
\end{aligned}
\end{equation}
The curvature is described in terms of the Ricci scalar $R$, components of the TF Ricci tensor $\Phi_{ij}$, and the Weyl scalars~$\Psi_k$, which are defined as:
\begin{equation}\label{eq:NPformalismCS}
\begin{aligned}
    \Phi_{00} &=\bar{\Phi}_{00}\equiv\tfrac12\bs{S_{ab}l^a l^b}\;,
    & 
    \Psi_0 &\equiv\bs{C_{abcd}l^a m^b l^c m^d}\;,
    \\
    \Phi_{01} &=\bar{\Phi}_{10}\equiv\tfrac12\bs{S_{ab}l^a m^b}\;,
    & 
    \Psi_1 &\equiv\bs{C_{abcd}l^a n^b l^c m^d}\;,
    \\
    \Phi_{02} &=\bar{\Phi}_{20}\equiv\tfrac12\bs{S_{ab}m^a m^b}\;,
    & 
    \Psi_2 &\equiv-\bs{C_{abcd}l^a m^b n^c} \bar{\bs{m}}\bs{{}^d}\;,
    \\
    \Phi_{11} &=\bar{\Phi}_{11}\equiv\tfrac14\bs{S_{ab}}\big(\bs{l^a n^b}+\bs{m^a} \bar{\bs{m}}\bs{{}^b}\big)\;,
    & 
    \Psi_3 &\equiv\bs{C_{abcd}n^a l^b n^c} \bar{\bs{m}}\bs{{}^d}\;,
    \\
    \Phi_{12} &=\bar{\Phi}_{21}\equiv\tfrac12\bs{S_{ab}n^a m^b}\;,
    &
    \Psi_4 &\equiv\bs{C_{abcd}n^a} \bar{\bs{m}}\bs{{}^b} \bs{n^c} \bar{\bs{m}}\bs{{}^d}\;,
    \\
    \Phi_{22} &=\bar{\Phi}_{22}\equiv\tfrac12\bs{S_{ab}n^a n^b}\;.
    &
\end{aligned}
\end{equation}
The components of the curvature and spin coefficients are connected through the Ricci identities:
\begin{equation}\label{eq:RicciIds}
\begin{aligned}
    \DD\rho-\bar{\delta} \kappa &= \rho^{2}+\sigma \bar{\sigma}+(\varepsilon+\bar{\varepsilon}) \rho-\bar{\kappa} \tau-\kappa(3 \alpha+\bar{\beta}-\pi)+\Phi_{00}\;,
    \\
    \DD\sigma-\delta \kappa &=(\rho+\bar{\rho}) \sigma+(3 \varepsilon-\bar{\varepsilon}) \sigma-(\tau-\bar{\pi}+\bar{\alpha}+3 \beta) \kappa+\Psi_{0}\;,
    \\
    \DD\tau-\nDelta \kappa &=(\tau+\bar{\pi}) \rho+(\bar{\tau}+\pi) \sigma+(\varepsilon-\bar{\varepsilon}) \tau-(3 \gamma+\bar{\gamma}) \kappa +\Psi_{1}+\Phi_{01}\;,
    \\
    \DD\alpha-\bar{\delta} \varepsilon &=(\rho+\bar{\varepsilon}-2 \varepsilon) \alpha+\beta \bar{\sigma}-\bar{\beta} \varepsilon-\kappa \lambda-\bar{\kappa} \gamma+(\varepsilon+\rho) \pi+\Phi_{10}\;,
    \\
    \DD\beta-\delta \varepsilon &=(\alpha+\pi) \sigma+(\bar{\rho}-\bar{\varepsilon}) \beta-(\mu+\gamma) \kappa-(\bar{\alpha}-\bar{\pi}) \varepsilon+\Psi_{1}\;,
    \\
    \DD\gamma-\nDelta \varepsilon &=(\tau+\bar{\pi}) \alpha+(\bar{\tau}+\pi) \beta-(\varepsilon+\bar{\varepsilon}) \gamma-(\gamma+\bar{\gamma}) \varepsilon +\tau \pi-\nu \kappa+\Psi_{2}+\Phi_{11}-R / 24\;,
    \\
    \DD\lambda-\bar{\delta} \pi &= \rho \lambda+\bar{\sigma} \mu+\pi^{2}+(\alpha-\bar{\beta}) \pi-\nu \bar{\kappa}-(3 \varepsilon-\bar{\varepsilon}) \lambda+\Phi_{20}\;,
    \\
    \DD\mu-\delta \pi &= \bar{\rho} \mu+\sigma \lambda+\pi \bar{\pi}-(\varepsilon+\bar{\varepsilon}) \mu-\pi(\bar{\alpha}-\beta)-\nu \kappa +\Psi_{2}+R / 12\;,
    \\
    \DD\nu-\nDelta \pi &=(\pi+\bar{\tau}) \mu+(\bar{\pi}+\tau) \lambda+(\gamma-\bar{\gamma}) \pi-(3 \varepsilon+\bar{\varepsilon}) \nu+\Psi_{3}+\Phi_{21}\;,
    \\
    \nDelta \lambda-\bar{\delta} \nu &=-(\mu+\bar{\mu}) \lambda-(3 \gamma-\bar{\gamma}) \lambda+(3 \alpha+\bar{\beta}+\pi-\bar{\tau}) \nu-\Psi_{4}\;,
    \\
    \delta \rho-\bar{\delta} \sigma &= \rho(\bar{\alpha}+\beta)-\sigma(3 \alpha-\bar{\beta})+(\rho-\bar{\rho}) \tau+(\mu-\bar{\mu}) \kappa -\Psi_{1}+\Phi_{01}\;,
    \\
    \delta \alpha-\bar{\delta} \beta &= \mu \rho-\lambda \sigma+\alpha \bar{\alpha}+\beta \bar{\beta}-2 \alpha \beta+\gamma(\rho-\bar{\rho})+\varepsilon(\mu-\bar{\mu}) -\Psi_{2}+\Phi_{11}+R / 24\;,
    \\
    \delta \lambda-\bar{\delta} \mu &=(\rho-\bar{\rho}) \nu+(\mu-\bar{\mu}) \pi+\mu(\alpha+\bar{\beta})+\lambda(\bar{\alpha}-3 \beta)-\Psi_{3}+\Phi_{21}\;,
    \\
    \delta \nu-\nDelta \mu &= \mu^{2}+\lambda \bar{\lambda}+(\gamma+\bar{\gamma}) \mu-\bar{\nu} \pi+(\tau-3 \beta-\bar{\alpha}) \nu+\Phi_{22}\;,
    \\
    \delta \gamma-\nDelta \beta &=(\tau-\bar{\alpha}-\beta) \gamma+\mu \tau-\sigma \nu-\varepsilon \bar{\nu}-\beta(\gamma-\bar{\gamma}-\mu) +\alpha \bar{\lambda}+\Phi_{12}\;,
    \\
    \delta \tau-\nDelta \sigma &= \mu \sigma+\bar{\lambda} \rho+(\tau+\beta-\bar{\alpha}) \tau-(3 \gamma-\bar{\gamma}) \sigma-\kappa \bar{\nu}+\Phi_{02}\;,
    \\
    \nDelta \rho-\bar{\delta} \tau &=-(\rho \bar{\mu}+\sigma \lambda)+(\bar{\beta}-\alpha-\bar{\tau}) \tau+(\gamma+\bar{\gamma}) \rho+\nu \kappa-\Psi_{2}-R / 12\;,
    \\
    \nDelta \alpha-\bar{\delta} \gamma &=(\rho+\varepsilon) \nu-(\tau+\beta) \lambda+(\bar{\gamma}-\bar{\mu}) \alpha+(\bar{\beta}-\bar{\tau}) \gamma-\Psi_{3}\;.
\end{aligned}
\end{equation}
Finally, it is also useful to list the equations that are equivalent to Bianchi identities in NP formalism:
\begin{equation}\label{eq:BianchiIds}
\begin{aligned}
    \bar{\delta} \Psi_{0} -\DD \Psi_{1}+\DD\Phi_{01}-\delta \Phi_{00} &=(4 \alpha-\pi) \Psi_{0}-2(2 \rho+\varepsilon) \Psi_{1}+3 \kappa \Psi_{2} +(\bar{\pi}-2 \bar{\alpha}-2 \beta) \Phi_{00}+2(\varepsilon+\bar{\rho}) \Phi_{01}
    \\
    &\feq+2 \sigma \Phi_{10}-2 \kappa \Phi_{11}-\bar{\kappa} \Phi_{02}\;,
    \\
    \nDelta \Psi_{0} -\delta \Psi_{1}+\DD\Phi_{02}-\delta \Phi_{01} &=(4 \gamma-\mu) \Psi_{0}-2(2 \tau+\beta) \Psi_{1}+3 \sigma \Psi_{2}+(2 \varepsilon-2 \bar{\varepsilon}+\bar{\rho}) \Phi_{02}+2(\bar{\pi}-\beta) \Phi_{01}
    \\
    &\feq+2 \sigma \Phi_{11}-2 \kappa \Phi_{12}-\bar{\lambda} \Phi_{00}\;,
    \\
    \bar{\delta} \Psi_{3} -\DD\Psi_{4}+\bar{\delta} \Phi_{21}-\nDelta \Phi_{20} &=(4 \varepsilon-\rho) \Psi_{4}-2(2 \pi+\alpha) \Psi_{3}+3 \lambda \Psi_{2}+(2 \gamma-2 \bar{\gamma}+\bar{\mu}) \Phi_{20}+2(\bar{\tau}-\alpha) \Phi_{21}
    \\
    &\feq+2 \lambda \Phi_{11}-2 \nu \Phi_{10}-\bar{\sigma} \Phi_{22}\;,
    \\
    \nDelta \Psi_{3} -\delta \Psi_{4}+\bar{\delta} \Phi_{22}-\nDelta \Phi_{21} &=(4 \beta-\tau) \Psi_{4}-2(2 \mu+\gamma) \Psi_{3}+3 \nu \Psi_{2}+(\bar{\tau}-2 \bar{\beta}-2 \alpha) \Phi_{22}+2(\gamma+\bar{\mu}) \Phi_{21}
    \\
    &\feq+2 \lambda \Phi_{12}-2 \nu \Phi_{11}-\bar{\nu} \Phi_{20}\;,
    \\
    \DD\Psi_{2} -\bar{\delta} \Psi_{1}+\nDelta \Phi_{00}-\bar{\delta} \Phi_{01}+\tfrac{1}{12} \DD R &=-\lambda \Psi_{0}+2(\pi-\alpha) \Psi_{1}+3 \rho \Psi_{2}-2 \kappa \Psi_{3}+(2 \gamma+2 \bar{\gamma}-\bar{\mu}) \Phi_{00}-2(\bar{\tau}+\alpha) \Phi_{01}
    \\
    &\feq-2 \tau \Phi_{10}+2 \rho \Phi_{11}+\bar{\sigma} \Phi_{02}\;,
    \\
    \nDelta \Psi_{2} -\delta \Psi_{3}+\DD\Phi_{22}-\delta \Phi_{21}+\tfrac{1}{12} \nDelta R &= \sigma \Psi_{4}+2(\beta-\tau) \Psi_{3}-3 \mu \Psi_{2}+2 \nu \Psi_{1}+(\bar{\rho}-2 \varepsilon-2 \bar{\varepsilon}) \Phi_{22}+2(\bar{\pi}+\beta) \Phi_{21}
    \\
    &\feq+2 \pi \Phi_{12}-2 \mu \Phi_{11}-\bar{\lambda} \Phi_{20}\;,
    \\
    \DD\Psi_{3} -\bar{\delta} \Psi_{2}-\DD\Phi_{21}+\delta \Phi_{20}-\tfrac{1}{12} \bar{\delta} R &=-\kappa \Psi_{4}+2(\rho-\varepsilon) \Psi_{3}+3 \pi \Psi_{2}-2 \lambda \Psi_{1}+(2 \bar{\alpha}-2 \beta-\bar{\pi}) \Phi_{20}-2(\bar{\rho}-\varepsilon) \Phi_{21}
    \\
    &\feq-2 \pi \Phi_{11}+2 \mu \Phi_{10}+\bar{\kappa} \Phi_{22}\;,
    \\
    \nDelta \Psi_{1}-\delta \Psi_{2}-\nDelta \Phi_{01}+\bar{\delta} \Phi_{02}-\tfrac{1}{12} \delta R &= \nu \Psi_{0}+2(\gamma-\mu) \Psi_{1}-3 \tau \Psi_{2}+2 \sigma \Psi_{3}+(\bar{\tau}-2 \bar{\beta}+2 \alpha) \Phi_{02}+2(\bar{\mu}-\gamma) \Phi_{01}
    \\
    &\feq+2 \tau \Phi_{11}-2 \rho \Phi_{12}-\bar{\nu} \Phi_{00}\;,
    \\
    \DD\Phi_{11} -\delta \Phi_{10}-\bar{\delta} \Phi_{01}+\nDelta \Phi_{00}+\tfrac{1}{8} \DD R &=(2 \gamma-\mu+2 \bar{\gamma}-\bar{\mu}) \Phi_{00}+(\pi-2 \alpha-2 \bar{\tau}) \Phi_{01}+(\bar{\pi}-2 \bar{\alpha}-2 \tau) \Phi_{10} 
    \\
    &\feq+2(\rho+\bar{\rho}) \Phi_{11}+\bar{\sigma} \Phi_{02}+\sigma \Phi_{20}-\bar{\kappa} \Phi_{12}-\kappa \Phi_{21}\;,
    \\
    \DD\Phi_{12} -\delta \Phi_{11}-\bar{\delta} \Phi_{02}+\nDelta \Phi_{01}+\tfrac{1}{8} \delta R &=(-2 \alpha+2 \bar{\beta}+\pi-\bar{\tau}) \Phi_{02}+(\bar{\rho}+2 \rho-2 \bar{\varepsilon}) \Phi_{12}+2(\bar{\pi}-\tau) \Phi_{11}
    \\
    &\feq+(2 \gamma-2 \bar{\mu}-\mu) \Phi_{01}+\bar{\nu} \Phi_{00}-\bar{\lambda} \Phi_{10}+\sigma \Phi_{21}-\kappa \Phi_{22}\;, 
    \\
    \DD\Phi_{22} -\delta \Phi_{21}-\bar{\delta} \Phi_{12}+\nDelta \Phi_{11}+\tfrac{1}{8} \nDelta R &=(\rho+\bar{\rho}-2 \varepsilon-2 \bar{\varepsilon}) \Phi_{22}+(2 \bar{\beta}+2 \pi-\bar{\tau}) \Phi_{12}+(2 \beta+2 \bar{\pi}-\tau) \Phi_{21}
    \\
    &\feq-2(\mu+\bar{\mu}) \Phi_{11}+\nu \Phi_{01}+\bar{\nu} \Phi_{10}-\bar{\lambda} \Phi_{20}-\lambda \Phi_{02}\;.
\end{aligned}
\end{equation}

\section{Explicit calculations for contractions of \texorpdfstring{$\bs{\nabla}{\cdots}\bs{\nabla} \bs{S}\bs{\nabla}{\cdots}\bs{\nabla}\bs{S}$}{}}\label{apx:explcalcSS}

This appendix contains a simple example of explicit calculations that were schematically indicated in Section~\ref{ssec:TensorsQuadraticInCurvature}. The schematic notation in \eqref{eq:SSterms} is equivalent to:
\begin{equation}
\begin{aligned}
    \bs{\nabla}{\cdots}\bs{\nabla} \bs{S^a{}_a}\bs{\nabla}{\cdots}\bs{\nabla}\bs{S} &=0\;,
    \\
    \bs{\nabla}{\cdots}\bs{\nabla}\bs{\nabla_a}\bs{\nabla}{\cdots}\bs{\nabla}\bs{\nabla_b}\bs{\nabla}{\cdots}\bs{\nabla} \bs{S^{ab}}\bs{\nabla}{\cdots}\bs{\nabla}\bs{S} &=\bs{\nabla}{\cdots}\bs{\nabla} \DD \bs{\nabla}{\cdots}\bs{\nabla}\bs{\nabla_b} \bs{\nabla}{\cdots}\bs{\nabla} (-2\Phi_{21} \bs{m^{b}}-2\bar{\Phi}_{21} \bs{\bar{m}^{b}}+\Phi_{22} \bs{ l^b})\bs{\nabla}{\cdots}\bs{\nabla}\bs{S}
    \\
    &+\bs{\nabla}{\cdots}\bs{\nabla} \bs{\nabla_a} \bs{\nabla}{\cdots}\bs{\nabla} \DD \bs{\nabla}{\cdots}\bs{\nabla} (-2\Phi_{21} \bs{m^{a}}-2\bar{\Phi}_{21} \bs{\bar{m}^{a}}+\Phi_{22} \bs{ l^a})\bs{\nabla}{\cdots}\bs{\nabla}\bs{S}=0\;,
    \\
    \bs{\nabla}{\cdots}\bs{\nabla}\bs{\nabla_a}\bs{\nabla}{\cdots}\bs{\nabla} \bs{S^{ab}}\bs{\nabla}{\cdots}\bs{\nabla}\bs{\nabla_b}\bs{\nabla}{\cdots}\bs{\nabla}\bs{S}
    &=\bs{\nabla}{\cdots}\bs{\nabla}\bs{\nabla_a}\bs{\nabla}{\cdots}\bs{\nabla} (-2\Phi_{21} \bs{m^{a}}-2\bar{\Phi}_{21} \bs{\bar{m}^{a}}+\Phi_{22} \bs{ l^a})\bs{\nabla}{\cdots}\bs{\nabla}\DD\bs{\nabla}{\cdots}\bs{\nabla}\bs{S}
    \\
    &+\bs{\nabla}{\cdots}\bs{\nabla}\DD\bs{\nabla}{\cdots}\bs{\nabla} (-2\Phi_{21} \bs{m^{b}}-2\bar{\Phi}_{21} \bs{\bar{m}^{b}}+\Phi_{22} \bs{ l^b})\bs{\nabla}{\cdots}\bs{\nabla}\bs{\nabla_b}\bs{\nabla}{\cdots}\bs{\nabla}\bs{S}=0\;,
    \\
    \bs{\nabla}{\cdots}\bs{\nabla} \bs{S^{ab}}\bs{\nabla}{\cdots}\bs{\nabla}\bs{\nabla_a}\bs{\nabla}{\cdots}\bs{\nabla}\bs{\nabla_b}\bs{\nabla}{\cdots}\bs{\nabla}\bs{S}
    &=\bs{\nabla}{\cdots}\bs{\nabla}(-2\Phi_{21} \bs{m^{b}}-2\bar{\Phi}_{21} \bs{\bar{m}^{b}}+\Phi_{22} \bs{ l^b})\bs{\nabla}{\cdots}\bs{\nabla}\DD\bs{\nabla}{\cdots}\bs{\nabla}\bs{\nabla_b}\bs{\nabla}{\cdots}\bs{\nabla}\bs{S}
    \\
    &+\bs{\nabla}{\cdots}\bs{\nabla}(-2\Phi_{21} \bs{m^{a}}-2\bar{\Phi}_{21} \bs{\bar{m}^{a}}+\Phi_{22} \bs{ l^a})\bs{\nabla}{\cdots}\bs{\nabla}\bs{\nabla_a}\bs{\nabla}{\cdots}\bs{\nabla}\DD\bs{\nabla}{\cdots}\bs{\nabla}\bs{S}=0\;,
    \\
    \bs{\nabla}{\cdots}\bs{\nabla}\bs{\nabla_a}\bs{\nabla}{\cdots}\bs{\nabla} \bs{S^{ab}}\bs{\nabla}{\cdots}\bs{\nabla}\bs{S_{b\circ}} &=\bs{\nabla}{\cdots}\bs{\nabla}\bs{\nabla_a}\bs{\nabla}{\cdots}\bs{\nabla}(-2\Phi_{21} \bs{m^{a}}-2\bar{\Phi}_{21} \bs{\bar{m}^{a}}+\Phi_{22} \bs{ l^a})\bs{\nabla}{\cdots}\bs{\nabla}(\bs{S}\cdot\bs{l}^{\sharp}) 
    \\
    &+ \bs{\nabla}{\cdots}\bs{\nabla} \DD\bs{\nabla}{\cdots}\bs{\nabla}(-2\Phi_{21} \bs{m^{b}}-2\bar{\Phi}_{21} \bs{\bar{m}^{b}}+\Phi_{22} \bs{ l^b})\bs{\nabla}{\cdots}\bs{\nabla}\bs{S_{b\circ}}=0\;,
    \\
    \bs{\nabla}{\cdots}\bs{\nabla} \bs{S^{ab}}\bs{\nabla}{\cdots}\bs{\nabla}\bs{\nabla_a}\bs{\nabla}{\cdots}\bs{\nabla}\bs{S_{b\circ}} &=\bs{\nabla}{\cdots}\bs{\nabla}(-2\Phi_{21} \bs{m^{a}}-2\bar{\Phi}_{21} \bs{\bar{m}^{a}}+\Phi_{22} \bs{ l^a})\bs{\nabla}{\cdots}\bs{\nabla}\bs{\nabla_a}\bs{\nabla}{\cdots}\bs{\nabla}(\bs{S}\cdot\bs{l}^{\sharp}) 
    \\
    &+ \bs{\nabla}{\cdots}\bs{\nabla}(-2\Phi_{21} \bs{m^{b}}-2\bar{\Phi}_{21} \bs{\bar{m}^{b}}+\Phi_{22} \bs{ l^b})\bs{\nabla}{\cdots}\bs{\nabla} \DD\bs{\nabla}{\cdots}\bs{\nabla}\bs{S_{b\circ}}=0\;,
    \\
    \bs{\nabla}{\cdots}\bs{\nabla} \bs{S^{ab}}\bs{\nabla}{\cdots}\bs{\nabla}\bs{S_{ab}} &=\bs{\nabla}{\cdots}\bs{\nabla}(-2\Phi_{21} \bs{m^{b}}-2\bar{\Phi}_{21} \bs{\bar{m}^{b}}+\Phi_{22} \bs{ l^b})\bs{\nabla}{\cdots}\bs{\nabla}(\bs{S_{ab}l^a})
    \\
    & +\bs{\nabla}{\cdots}\bs{\nabla}(-2\Phi_{21} \bs{m^{a}}-2\bar{\Phi}_{21} \bs{\bar{m}^{a}}+\Phi_{22} \bs{ l^a})\bs{\nabla}{\cdots}\bs{\nabla}(\bs{S_{ab}l^b})=0\;.
\end{aligned}    
\end{equation}


\section{Auxiliary statements for pp-waves of type III}\label{apx:auxstattypeIIIpp}

In this appendix, we list some rank-2 tensors linear or quadratic in curvature and point out their properties in type III pp-wave spacetimes that are used in the main text:

\begin{itemize}
    \item $\square^n \bs{S}$ is divergence-free.
        \begin{equation}\label{eq:divboxS}
            \bs{\nabla_b} \square^n \bs{S^{ab}} = 0\;.
        \end{equation}
        This property of $\square^n \bs{S}$ can be proved by induction. The case ${n=0}$ follows directly from the contracted Bianchi identities \eqref{eq:contractedBianchi}. Now, let us assume that \eqref{eq:divboxS} holds for $n$ and show that it then also holds for ${n+1}$ by commuting $\bs{\nabla}$ over one $\square$, i.e.\
        \begin{equation}
        \begin{aligned}
            \bs{\nabla_b} \square^{n+1} \bs{S^{ab}} &= \square \bs{\nabla_b} \square^n \bs{S^{ab}} - \bs{S^{bc} \nabla^{a}}\square^n \bs{S_{bc}} - \tfrac12 \square^n \bs{S^{bc} \nabla^a S_{bc}} + 2 \bs{S^{bc} \nabla_b} \square^k \bs{S^{a}{}_{c}} + \bs{S^{ab} \nabla^c} \square^n \bs{S_{bc}} \\
            &\feq + \tfrac12 \square^n \bs{S^{bc} \nabla_b S^{a}{}_{c}} + \tfrac32 \square^n \bs{S^{ab} \nabla^c S_{bc}} + \square^n \bs{S^{bc} \nabla^d C^{a}{}_{bcd}} + 2 \bs{C^{a}{}_{bcd} \nabla^d} \square^n \bs{S^{bc}} = 0\;,
        \end{aligned}
        \end{equation}
        where we employed the assumption \eqref{eq:divboxS} for $n$ and the results of Section \ref{ssec:TensorsQuadraticInCurvature}.
        
    \item $\bs{C_{acbd}} \square^n \bs{S^{cd}}$, $\bs{S^{cd}} \square^n \bs{C_{acbd}}$, and $\bs{S_a{}^c} \square^n \bs{S_{bc}}$ are equal. \\
        Rank-2 tensors constructed from two tensors of b.o.\ $-1$ are of b.o.\ $-2$, i.e., they have only b.w.\ $-2$ parts proportional to $\bs{l_a l_b}$ and therefore only the b.w.\ $-1$ parts of two original tensors contribute. As mentioned in Section \ref{ssec:typeIIIppwaves}, only $\bs{S}^\bwpart{-1}$ or $\bs{C}^\bwpart{-1}$ and their covariant derivatives give the b.w.\ $-1$ parts of $\bs{\nabla}{\cdots}\bs{\nabla S}$ or $\bs{\nabla}{\cdots}\bs{\nabla C}$ of type III pp-waves, respectively. Hence,
        \begin{equation}
        \begin{aligned}
            \bs{C_{acbd}} \square^k \bs{S^{cd}} 
            &= 8 \bs{l_a l_b} [\Psi_3 \bs{m}_{\bs{(c}} \bs{n}_{\bs{d)}} + \bar{\Psi}_3 \bar{\bs{m}}_{\bs{(c}} \bs{n}_{\bs{d)}}] \square^k [\Phi_{21} \bs{l}^{\bs{(c}} \bs{m}^{\bs{d)}} + \bar{\Phi}_{21} \bs{l}^{\bs{(c}} \bar{\bs{m}}^{\bs{d)}}]
            \\
            &=- 4 \bs{l_a l_b} [\Psi_3 \bs{m}_{\bs{c}} + \bar{\Psi}_3 \bar{\bs{m}}_{\bs{c}}] \square^k [\Phi_{21} \bs{m}^{\bs{c}} + \bar{\Phi}_{21} \bar{\bs{m}}^{\bs{c}}]\;, \\
            \bs{S^{cd}} \square^k \bs{C_{acbd}} 
            &= 8 \bs{l_a l_b} [\Phi_{21} \bs{l}^{\bs{(c}} \bs{m}^{\bs{d)}} + \bar{\Phi}_{21} \bs{l}^{\bs{(c}} \bar{\bs{m}}^{\bs{d)}}] \square^k [\Psi_3 \bs{m}_{(\bs{c}} \bs{n}_{\bs{d})} + \bar{\Psi}_3 \bar{\bs{m}}_{\bs{(c}} \bs{n}_{\bs{d)}}]
            \\
            &=- 4 \bs{l_a l_b} [\Phi_{21} \bs{m}^{\bs{c}} + \bar{\Phi}_{21} \bar{\bs{m}}^{\bs{c}}] \square^k [\Psi_3 \bs{m}_{\bs{c}} + \bar{\Psi}_3 \bar{\bs{m}}_{\bs{c}}]\;, \\
            \bs{S_a{}^c} \square^k \bs{S_{bc}} &= 4 [\Phi_{21} \bs{l_a m^c} + \bar{\Phi}_{21} \bs{l_a \bar{m}^c}] \square^k [\Phi_{21} \bs{l_b m_c} + \bar{\Phi}_{21} \bs{l_b \bar{m}_c}] \\
            &= 4 \bs{l_a l_b} [\Phi_{21} \bs{m^c} + \bar{\Phi}_{21} \bs{\bar{m}^c}] \square^k [\Phi_{21} \bs{m_c} + \bar{\Phi}_{21} \bs{\bar{m}_c}]\;.
        \end{aligned}
        \end{equation}
        The equality of the terms is then a consequence of ${\Psi_3 = - \Phi_{21}}$,
        \begin{equation}\label{eq:SboxS}
            \bs{C_{acbd}} \square^k \bs{S^{cd}} = \bs{S^{cd}} \square^k \bs{C_{acbd}} = \bs{S_a{}^c} \square^k \bs{S_{bc}}\;.
        \end{equation}
        
    \item $\bs{\nabla_c \nabla_d} \square^k \bs{C_{ab}{}^{cd}}$ vanishes. \\
        Using the cyclic symmetry of the Weyl tensor ${\bs{\nabla_c \nabla_d} \square^k \bs{C_a}{}^{[\bs{bcd}]} = 0}$ and the commutator of $\bs{\nabla_c\nabla_d}$, one can show that
        \begin{equation}\label{eq:NablacNabladBoxCabcd}
            \bs{\nabla_c \nabla_d} \square^k \bs{C_{ab}{}^{cd}} = - \bs{C_{[a}{}^{cde}} \square^k \bs{C_{b]cde}}\;.
        \end{equation}
        The right-hand side vanishes because the rank-2 tensor $\bs{C_{a}{}^{cde}} \square^k \bs{C_{bcde}}$ is of b.o.\ $-2$ for type III pp-waves and therefore proportional to $\bs{l_a l_b}$.
        
    \item $\bs{\nabla_d} \square^k \bs{C_{acbe} \nabla^e S^{cd}}$ and
    $\bs{\nabla_e} \square^k \bs{C_{acbd} \nabla^e S^{cd}}$ equal
    $\frac12\bs{\nabla_{c}S_{da} \nabla^d} \square^k \bs{S_{b}{}^{c}}$ and
    $\bs{\nabla_d S_{ca} \nabla^d} \square^k \bs{\nabla S_b{}^c}$, respectively.
    
        Straightforwardly from the decomposition of the TF Ricci and Weyl tensors of type III pp-waves \eqref{eq:SCtypeIII} along with $\DD\bs{\nabla}{\cdots}\bs{\nabla}C = 0$ due to \eqref{eq:DnablaR}, it follows that
        \begin{equation}
        \begin{aligned}
            \bs{\nabla_{c}S_{da} \nabla^d} \square^k \bs{S_{b}{}^{c}} &= 4 \bs{l_a l_b \nabla_c}[\Phi_{21} \bs{m_d} + \bar{\Phi}_{21} \bs{\bar{m}_d}] \bs{\nabla^d} \square^k[\Phi_{21} \bs{m^c} + \bar{\Phi}_{21} \bs{\bar{m}^c}]\;, \\
            \bs{\nabla_d} \square^k \bs{C_{acbe} \nabla^e S^{cd}} 
            &= 4 \bs{l_a l_b \nabla_d} \square^k [\Psi_3 \bs{m_{(c} n_{e)}} + \bar\Psi_3 \bs{\bar{m}_{(c} n_{e)}}] \bs{\nabla^e}[\Phi_{21} \bs{l^c m^d} + \bar{\Phi}_{21} \bs{l^c \bar{m}^d}] \\
            &= -2 \bs{l_a l_b \nabla_d} \square^k [\Psi_3 \bs{m_e} + \bar\Psi_3 \bs{\bar{m}_e}] \bs{\nabla^e}[\Phi_{21} \bs{m^d} + \bar{\Phi}_{21} \bs{\bar{m}^d}]\;, \\
            \bs{\nabla_d S_{ca} \nabla^d} \square^k \bs{S_b{}^c} &= 4 \bs{l_a l_b \nabla_d}[\Phi_{21} \bs{m_c} + \bar{\Phi}_{21} \bs{\bar{m}_c}] \bs{\nabla^d} \square^k[\Phi_{21} \bs{m^c} + \bar{\Phi}_{21} \bs{\bar{m}^c}]\;, \\
            \bs{\nabla_e} \square^k \bs{C_{acbd} \nabla^e S^{cd}} &= 8 \bs{l_a l_b \nabla_e} \square^k [\Psi_3 \bs{m_{(c} n_{d)}} + \bar\Psi_3 \bs{\bar{m}_{(c} n_{d)}}] \bs{\nabla^e} [\Phi_{21} \bs{l^{(c} m^{d)}} + \bar\Phi_{21} \bs{l^{(c} \bar{m}^{d)}}] \\
            &= -4 \bs{l_a l_b \nabla_e} \square^k [\Psi_3 \bs{m_c} + \bar\Psi_3 \bs{\bar{m}_c}] \bs{\nabla^e} [\Phi_{21} \bs{m^c} + \bar\Phi_{21} \bs{\bar{m}^c}]\;.
        \end{aligned}
        \end{equation}
        Substituting $\Psi_3 = - \Phi_{21}$, we obtain
        \begin{equation}\label{eq:greenorangeterms}
            \bs{\nabla_d} \square^k \bs{C_{acbe} \nabla^e S^{cd}} = \tfrac12 \bs{\nabla_{c}S_{da} \nabla^d} \square^k \bs{S_{b}{}^{c}},
            \quad \bs{\nabla_e} \square^k \bs{C_{acbd} \nabla^e S^{cd}} = \bs{\nabla_d S_{ca} \nabla^d} \square^k \bs{S_b{}^c}\;.
        \end{equation}

    \item $\bs{C^{cdef} \nabla_d \nabla_f} \square^k \bs{C_{c(ab)e}}$ equals $\bs{\nabla^e\nabla^d}\square^k \bs{S_{(a}{}^c C_{b)dce}}$.

        If any index of $\bs{d}$, $\bs{c}$, or $\bs{e}$ in $\bs{C_{bdce}}$ is associated with $\bs{l}$, then $\bs{\nabla^e\nabla^d}\square^k \bs{S_{a}{}^c C_{bdce}}$ vanishes because $\DD\bs{\nabla}{\cdots}\bs{\nabla S} = \bs{S}\cdot\bs{l}^\sharp = 0$. Therefore, only the terms proportional to $\bs{l\bar{m}m\bar{m}}$ or $\bs{lm\bar{m}m}$ of the b.w.\ $-1$ part of $\bs{C_{bdce}}$ contribute,
        \begin{equation}\label{eq:cyanterm1}
            \bs{\nabla^e\nabla^d}\square^k \bs{S_{(a}{}^c C_{b)dce}} 
            = -4 \bs{l_a l_b \nabla^e\nabla^d}\square^k[\Phi_{21} \bs{m^c} + \bar\Phi_{21} \bs{\bar{m}^c}][\Psi_3 \bs{m_d \bar{m}_{[c} m_{e]}} + \bar\Psi_3 \bs{\bar{m}_d m_{[c} \bar{m}_{e]}}]\;.
        \end{equation}
        If index $\bs{d}$ or $\bs{f}$ of $\bs{C^{cdef}}$ is associated with $\bs{l}^\sharp$, then $\bs{C^{cdef} \nabla_d \nabla_f} \square^k \bs{C_{c(ab)e}}$ vanishes due to $\DD\bs{\nabla}{\cdots}\bs{\nabla C} = 0$. If both $\bs{c}$ and $\bs{e}$ are associated with $\bs{l}^\sharp$, this term vanishes as well due to ${\bs{l}^{\sharp} \cdot \bs{C} \cdot \bs{l}^{\sharp} = 0}$. Hence, only the terms proportional to $\bs{l}^\sharp\bs{\bar{m}}^\sharp \bs{m}^\sharp \bs{\bar{m}}^\sharp$ or $\bs{l}^\sharp \bs{m}^\sharp \bs{\bar{m}}^\sharp \bs{m}^\sharp$ of the b.w.\ $-1$ part of $\bs{C^{cdef}}$ contribute when $\bs{l}^\sharp$ is associated either with $\bs{c}$ or $\bs{e}$, i.e.,
        \begin{equation}\label{eq:cyanterm2}
        \begin{aligned}
            \bs{C^{cdef} \nabla_d \nabla_f} \square^k \bs{C_{c(ab)e}} 
            &= -2 \bs{l_a l_b} [\Psi_3 \bs{l^c m^d \bar{m}^{[e} m^{f]}} + \bar\Psi_3 \bs{l^c \bar{m}^d m^{[e} \bar{m}^{f]}}] \bs{\nabla_d\nabla_f} \square^k [\Psi_3 \bs{n_c m_e} + \bar\Psi_3 \bs{n_c \bar{m}_e}] \\
            &\quad -2 \bs{l_a l_b} [\Psi_3 \bs{l^e m^f \bar{m}^{[c} m^{d]}} + \bar\Psi_3 \bs{l^e \bar{m}^f m^{[c} \bar{m}^{d]}}] \bs{\nabla_d\nabla_f} \square^k [\Psi_3 \bs{m_c n_e} + \bar\Psi_3 \bs{\bar{m}_c n_e}] \\
            &= 2 \bs{l_a l_b} [\Psi_3 \bs{m^d \bar{m}^{[e} m^{f]}} + \bar\Psi_3 \bs{\bar{m}^d m^{[e} \bar{m}^{f]}}] \bs{\nabla_d\nabla_f} \square^k [\Psi_3 \bs{m_e} + \bar\Psi_3 \bs{\bar{m}_e}] \\
            &\quad 2 \bs{l_a l_b} [\Psi_3 \bs{m^f \bar{m}^{[c} m^{d]}} + \bar\Psi_3 \bs{\bar{m}^f m^{[c} \bar{m}^{d]}}] \bs{\nabla_d\nabla_f} \square^k [\Psi_3 \bs{m_c} + \bar\Psi_3 \bs{\bar{m}_c}] \\
            &= 4 \bs{l_a l_b} [\Psi_3 \bs{m^d \bar{m}^{[e} m^{f]}} + \bar\Psi_3 \bs{\bar{m}^d m^{[e} \bar{m}^{f]}}] \bs{\nabla_d\nabla_f} \square^k [\Psi_3 \bs{m_e} + \bar\Psi_3 \bs{\bar{m}_e}]\;,
        \end{aligned}
        \end{equation}
        where we employed the fact that, for type III pp-waves, the covariant derivatives commute in rank-2 tensors quadratic in curvature. Finally, comparing \eqref{eq:cyanterm1} with \eqref{eq:cyanterm2}, we get
        \begin{equation}\label{eq:cyanterms}
            \bs{C^{cdef} \nabla_d \nabla_f} \square^k \bs{C_{c(ab)e}} = \bs{\nabla^e\nabla^d} \square^k \bs{S_{(a}{}^c C_{b)dce}}\;.
        \end{equation}
\end{itemize}



\bibliography{references.bib}

\end{document}